\documentclass[aps,preprint,nofootinbib]{revtex4}
\usepackage[]{fontenc}
\usepackage[latin1]{inputenc}
\usepackage{amsmath}
\usepackage{graphicx}

\makeatletter

\providecommand{\tabularnewline}{\\}

\usepackage[english]{babel}
\makeatother
\begin{document}
\newcommand{\abs}[1]{\left| #1 \right| }
\newcommand{\half}{\frac{1}{2}}

\newcommand{\db}{\!\not\!\! D\,}

\newcommand{\mnu}{\mathcal{M}_{\nu}}
\newcommand{\phd}{\delta}
\newcommand{\phm}{\phi}
\newcommand{\fem}{f_{e\mu}}
\newcommand{\fet}{f_{e\tau}}
\newcommand{\fmt}{f_{\mu\tau}}
\newcommand{\gee}{g_{ee}}
\newcommand{\gem}{g_{e\mu}}
\newcommand{\get}{g_{e\tau}}
\newcommand{\gmm}{g_{\mu\mu}}
\newcommand{\gmt}{g_{\mu\tau}}
\newcommand{\gtt}{g_{\tau\tau}}

\preprint{FTUV/07-1103, IFIC/07-70, RM3-TH/07-16\\
}

\title{Prospects for the Zee-Babu Model at the LHC\\
 and low energy experiments}

\author{Miguel Nebot}

\affiliation{Centro de Física Teórica de Partículas (CFTP), \\
Instituto Superior Técnico, P-1049-001, Lisboa, Portugal}

\affiliation{Instituto Nazionale di Fisica Nucleare (INFN), Sezione di Roma Tre,\\
 Dipartamento di Fisica {}``Edoardo Amaldi'', Università degli Studi
Roma Tre, I-00146, Roma, Italy.}

\author{Josep F. Oliver, David Palao and Arcadi Santamaria}

\affiliation{Departament de Física Teòrica, Universitat de València and\\
IFIC, Universitat de València-CSIC\\
Dr. Moliner 50, E-46100 Burjassot (València), Spain}

\begin{abstract}
We analyze the viability of the Zee-Babu model as an explanation of
observed neutrino masses and mixings and the possibility that the
model is confirmed or discarded in experiments planned for the very
close future. The allowed parameter space is studied analytically
by using some approximations and partial data. Then, a complete scanning
of all parameters and constraints is performed numerically by using
Monte Carlo methods. The cleanest signal of the model will be the
detection of the doubly charged scalar at the LHC and its correlation
with measurements of the branching ratio of $\mu\rightarrow e\gamma$
at the MEG experiment. In addition, the model offers interesting predictions
for $\tau^{-}\rightarrow\mu^{+}\mu^{-}\mu^{-}$ experiments, lepton-hadron
universality tests, the $\theta_{13}$ mixing in neutrino oscillations
and the $\langle m_{\nu}\rangle_{ee}$ parameter of neutrinoless double
beta decay. 
\end{abstract}
\maketitle

\section{introduction}

The hints for neutrino masses accumulated in the last 30 years \cite{Davis:1968cp,Hirata:1988uy}
have been converted into a strong evidence in the last 10 years \cite{Cleveland:1998nv,Hampel:1998xg,Altmann:2000ft,Ahmad:2002jz,Eguchi:2002dm,Fukuda:1998mi,Allison:1999ms,Aliu:2004sq}:
the only consistent explanation for solar neutrino data, atmospheric
neutrino data, reactor and accelerator neutrino experiments is based
on the hypothesis of massive neutrinos which mix and oscillate. Using
this hypothesis all neutrino data can be fitted with just two squared
mass differences and three mixing angles. At present, phases are not
needed to explain the data although they can be included in the analyses
and hopefully will be tested in the future. 

This picture, although certainly a big step forward in our understanding
of neutrino physics leaves more open questions than the standard model
(SM) picture in which neutrinos are exactly massless because (1) the
SM does not contain righthanded neutrinos, thus, Dirac neutrinos are
not possible, and (2) the renormalizability of the model, with the
minimal Higgs content, enforces the exact conservation of the lepton
number%
\footnote{Perturbatively both lepton ($L$) and baryon ($B$) numbers are conserved
separately. If nonperturbative effects are taken into account only
 $B-L$ is exactly conserved. Still, $L$ and $B$ violations are
tiny at zero temperature and density. %
} which prevents Majorana mass terms. Therefore, to include neutrino
masses, we should relax some or several of the above ingredients of
the SM:

\begin{enumerate}
\item Add righthanded neutrinos.
\item Add new fields which could allow for lepton number violation while
keeping the renormalizability of the model.
\item Drop the renormalizability of the model.
\end{enumerate}
The last possibility is very general and allows for neutrino masses
without touching the field content of the SM. However, its predictivity
is very limited and it is only useful when seen as a low energy parametrization
of a more complete theory containing heavy non-standard particles.

Allowing for righthanded neutrinos seems the simplest and most economical
solution: if neutrinos are like all the other fermions, it is natural
to consider righthanded neutrinos coupled to the lefthanded neutrinos
and the Higgs scalar, obtaining in this way a Dirac mass term as all
the other fermions have. One can then easily adjust these Yukawa couplings
and fit all neutrino masses and mixings without problem. 

However, neutrinos are not like all the other fermions: their masses
are extremely small $m_{\nu}<1$~eV. In addition, righthanded neutrinos
are completely neutral with respect to the SM gauge group, therefore,
nothing forbids a righthanded Majorana mass term in the SM. In fact,
many will argue that it should be written unless a good reason is
given to forbid it. If this mass is present and if it is very large,
one can naturally explain the smallness of neutrino masses with the
seesaw mechanism and perhaps provide a way to understand the amount
of matter in the universe through the leptogenesis mechanism. Because
of all these virtues, this model, SM plus heavy righthanded neutrinos,
has almost universally become the standard mechanism for neutrino
masses. However, the model is not completely satisfactory: the heavy
Majorana neutrinos contribute, through loop effects, to the Higgs
mass with terms which are proportional to the square of the righthanded
Majorana mass. If this mass is huge, the contribution to the Higgs
mass will be too large. This is a specific realization of the generic
hierarchy problem of the SM. However, while within the strict SM one
could argue that the new heavy particles contributing to the Higgs
mass could have very small couplings to the Higgs boson, in the seesaw
mechanism we have explicit particles with explicit masses and couplings
giving contributions to the Higgs boson mass and the hierarchy problem
cannot be swept under the rug anymore. Thus, supersymmetry or any
other mechanism must be invoked to stabilize the Higgs boson mass.
On the other hand, the fact that the righthanded neutrino mass is
very large makes the effects of righthanded neutrinos negligible at
low energies, except, of course, those related with neutrino oscillations.
The model just provides neutrino masses and cannot be tested in present
or planned experiments%
\footnote{Although one can check it indirectly through its effects in leptogenesis.%
}.

Adding other fields to the SM gives many possibilities but obviously,
if there are no righthanded neutrinos, all of them require the non-conservation
of the lepton number and neutrinos acquire a Majorana mass term. Since
supersymmetry provides the best solution to the hierarchy problem,
one natural choice is to use the fields already present in the supersymmetric
extensions of the SM to generate neutrino masses. Very interesting
models of this type with spontaneous breakdown of R-parity (see for
instance \cite{Ellis:1984gi,Santamaria:1987uq}), which implies lepton
number violation, or explicit R-parity violation (see \cite{Hall:1983id,Hirsch:2000ef})
have been built. All these models basically use the sneutrinos to
generate neutrino masses. 

Alternatively one could add new ad hoc scalars to the SM in such a
way that lepton number is not automatically conserved: triplets which
develop a VEV \cite{Gelmini:1980re,Georgi:1981pg}, charged singlets
which could allow for explicit lepton number violation \cite{Zee:1980ai,Zee:1985id,Zee:1985rj,Babu:1988ki}
or a mixture of the two mechanisms (see for instance \cite{Bertolini:1987kz,Santamaria:1988fh}).

The seesaw mechanism is probably the most natural mechanism for neutrino
masses. However, the LHC is going to provide results very soon and
there are plans to increase the precision in $\mu\rightarrow e\gamma$,
$\mu\rightarrow eee$ experiments in a couple of orders of magnitude
(see for instance \cite{Ritt:2006cg}). It may be time to explore
alternatives for neutrino mass generation that could be confirmed
or rejected in planned experiments; among them, the natural candidates
are models in which neutrino masses are generated through radiative
corrections: as the masses will be suppressed by loop effects the
new particles responsible for them could be relatively light and be
produced at the LHC, at the International Linear Collider (ILC) and
may have sizable effects in $\mu\rightarrow e\gamma$, $\mu\rightarrow eee$
experiments. The simplest model of neutrino masses is the Zee-Babu\footnote{The model was first proposed in \cite{Zee:1985id,Zee:1985rj} and studied later in 
\cite{Babu:1988ki}. In the literature, it has
been often referred to as the Babu model.}  (ZB) 
model \cite{Zee:1985id,Zee:1985rj,Babu:1988ki} 
which just adds two complex
singlet scalar fields to the SM (that is, just 4 new degrees of freedom)
with neutrino masses generated at the two-loop level. Another very
interesting model is the Zee model \cite{Zee:1980ai} which adds a
new scalar doublet and a complex scalar singlet (6 new degrees of
freedom), however, the simplest version of the model gives a too sharp
bimaximal prediction for neutrino mixing and has already been excluded
\cite{Koide:2001xy,Balaji:2001ex,Oliver:2001eg}. Therefore, we will
only consider here the Zee-Babu model. In the Zee-Babu model neutrino
masses are generated at two loops and are proportional to several
Yukawa couplings of the new scalars and inversely proportional to
the square of their masses, therefore the couplings cannot be too
small and the scalar masses cannot be too large otherwise the generated
neutrino masses would be too small. This is very interesting because
the new scalars may be accessible at the LHC and could mediate the
processes $\mu\rightarrow e\gamma$ and $\mu\rightarrow eee$ with
rates measurable in planned experiments. In this paper we will sharpen
the predictions of the model by using both analytical and numerical
methods, specially under the assumption that the new scalars are light
enough to be produced at the LHC. The phenomenology of the Zee-Babu
model was recently reviewed in \cite{Babu:2002uu}%
\footnote{See also ref.~\cite{AristizabalSierra:2006gb}.%
} where under certain assumptions analytic limits on several couplings
and masses were set. The analysis of \cite{Babu:2002uu} is very clever
and interesting, however, the calculation of several processes in
that paper was taken from older papers with some wrong factors of
$2$ which, unfortunately, have also propagated to more recent papers.
In addition, ref.~\cite{Babu:2002uu} makes the simplifying assumption
that certain couplings are negligible. Thus, we found interesting
to review the phenomenology of the model by relaxing this assumption.
This, unavoidably requires a numerical study which will be presented
in this paper. We also take into account new stronger limits on flavour
lepton number violating tau decays from BELLE \cite{Abe:2006sf,Yusa:2004gm}
and BABAR \cite{Wilson:2006yu}. Thus, in section \ref{sec:The-Babu-model}
we sketch the model and review some of its features: the neutrino
masses, contributions to low energy processes like $\mu\rightarrow e\gamma$,
$\mu\rightarrow3e$, $\tau\rightarrow3\mu$ and favoured values for
the parameters of the model. In section \ref{sec:LHC} we study the
production and decays of the new scalars at the LHC. In section \ref{sec:Analysis}
we analyze all the relevant constraints on the parameters of the model
and obtain its predictions for lepton flavour violating processes
and for neutrino mass parameters ($\theta_{13}$ mixing, Majorana
and Dirac phases, neutrinoless double beta decay parameter $\langle m_{\nu}\rangle_{ee}$,
...). Section \ref{sec:Results} is devoted to a summary of the results.
Finally in section \ref{sec:Conclusions} we present our conclusions.

\section{The Zee-Babu model\label{sec:The-Babu-model}}

The Zee-Babu model is the minimal extension of the SM providing neutrino
masses and mixings compatible with experiment: in addition to the
Standard Model field spectrum, it only contains one singly charged
scalar and one doubly charged scalar.

In order to fix the notation we will briefly review the Zee-Babu model.
We will denote the SM particle content as follows: $\ell$ will be
the lefthanded lepton doublet, $\widetilde{\ell}\equiv i\tau_{2}\ell^{c}=i\tau_{2}C\bar{\ell}^{T}$
is just the conjugate lepton doublet used to build Majorana type couplings,
$e$ is the righthanded lepton and $H$ is the Higgs boson doublet.
We also have, of course, the weak gauge bosons $\vec{W}_{\mu}$, $B_{\mu}$,
and the quarks and gluons. As mentioned, the Zee-Babu model contains,
in addition, two charged singlet scalar fields\begin{equation}
h^{\pm},\qquad k^{\pm\pm}\,,\end{equation}
with weak hypercharges $\pm1$ and $\pm2$ respectively. Here we will
follow the convention $Q=T_{3}+Y$ and that $h$ and $k$ destroy
negatively charged particles, thus $h=h^{-}$ and $h^{\dagger}=h^{+}$,
while $k=k^{--}$ and $k^{\dagger}=k^{++}$. Then, the Lagrangian
can be split into two parts: \begin{equation}
\mathcal{L}=\mathcal{L}_{\mathrm{SM}}+\mathcal{L}_{\mathrm{ZB}}\,\,.\label{eq:lagrangian}\end{equation}
 The first part, $\mathcal{L}_{\mathrm{SM}}$, is the Standard Model
Lagrangian: \begin{equation}
\mathcal{L}_{\mathrm{SM}}=i\overline{\ell}\db\ell+i\overline{e}\db e+(\overline{\ell}Ye\, H+\mathrm{h.c.})+\cdots\label{eq:SMlagrangian}\end{equation}
The dots represent SM gauge boson, Higgs boson and quark kinetic terms,
quark Yukawa interactions and the SM Higgs potential. Generation and
$SU(2)$ indices have been suppressed, and therefore, $Y$ is a completely
general $3\times3$ matrix in generation space. The new terms in the
Lagrangian are

\begin{equation}
\mathcal{L}_{ZB}=D_{\mu}h^{\dagger}D^{\mu}h+D_{\mu}k^{\dagger}D^{\mu}k+\overline{\tilde{\ell}}f\ell h^{+}+\overline{e^{c}}g\, e\, k^{++}+\mathrm{h.c.}-V_{ZB}\,.\label{eq:babuyuk}\end{equation}
Since both $h$ and $k$ are $SU(2)$ singlets the covariant derivative
only contains couplings to the $B$ gauge boson, which after diagonalization
will generate photon and $Z$-boson couplings with the scalars, but
no $W$ couplings. Due to Fermi statistics the Zee type Yukawa coupling,
$f_{ab}$, is an antisymmetric matrix in flavour space while the Yukawa
coupling, $g_{ab}$, of the doubly charged scalar $k$, is a symmetric
matrix. The scalar potential $V_{ZB}$ contains all renormalizable
interactions between the scalars $h$, $k$ and between them and the
standard Higgs doublet: \begin{eqnarray}
V_{ZB} & = & m_{h}^{\prime2}|h|^{2}+m_{k}^{\prime2}|k|^{2}+\lambda_{h}|h|^{4}+\lambda_{k}|k|^{4}+\lambda_{hk}|h|^{2}|k|^{2}\nonumber \\
 & + & \lambda_{hH}|h|^{2}H^{\dagger}H+\lambda_{kH}|k|^{2}H^{\dagger}H+\left(\mu h^{2}k^{++}+\mathrm{h.c.}\right)\,.\end{eqnarray}
Trilinear terms, like in the Zee model, with two scalar doublets and
the $h$ vanish identically because the coupling is antisymmetric
in $SU(2)$ indices and, therefore, requires two different doublets.
The last term is particularly interesting because if $\mu\rightarrow0$,
the complete Lagrangian has an additional global $U(1)$ symmetry
which can be identified with lepton number $L$ (or $B-L$). In fact,
if we assign lepton number $1$ to both, the lepton doublet and the
righthanded lepton singlet, we can also assign lepton number $-2$
to both, the scalars $h$ and $k$, in such a way that this quantum
number is conserved in all the Lagrangian except in the trilinear
coupling of the scalar potential. Thus, if $\mu\not=0$, lepton number
is explicitly broken by the $\mu$-coupling in the scalar potential.
This is very important because this lepton number violation will be
transmitted to the fermionic sector and will finally be responsible
for the generation of neutrino masses. It is also important to remark
that this mechanism for lepton number violation requires the simultaneous
presence of the four couplings $Y$, $f$, $g$ and $\mu$, because
if any of them vanishes one can always assign quantum numbers in such
a way that there is a global $U(1)$ symmetry. This means that neutrino
masses will require the simultaneous presence of the four couplings. 

It is also important to note that the gauge-kinetic part of the Lagrangian
is invariant under the following $U(N)$ transformations in generation
space (for $N$ generations of leptons). \begin{equation}
\ell\rightarrow U_{\ell}\ell\:,\qquad e\rightarrow U_{e}e\,.\label{eq:FlavourTransformations}\end{equation}
Yukawa couplings, however, break this symmetry. This implies that
sets of Yukawa couplings related by the following redefinitions  are
completely equivalent\begin{equation}
(Y,f,g)\rightarrow(U_{\ell}^{\dagger}YU_{e},U_{\ell}^{T}fU_{\ell},U_{e}^{T}gU_{e})\,,\label{eq:redefinitions}\end{equation}
which in turn means that physical observables should transform correctly
under these redefinitions. This is important to check the behaviour
under flavour transformations of physical amplitudes, moreover it
can also be used to choose a convenient set of parameters in the Yukawa
sector and count the number of physical parameters following the methods
developed in \cite{Santamaria:1993ah}. Thus, using redefinitions
of eq.~\eqref{eq:redefinitions} one can choose, without loss of
generality, $Y$ diagonal with real and positive elements. One could
also choose $f$ real and antisymmetric and leave $g$ as a completely
general complex symmetric matrix. In addition, one can use redefinitions
of $h^{+}$ and $k^{++}$ to set $\mu$ real and positive and to remove
one of the phases in $g$. Thus, we finally have%
\footnote{The counting can be generalized to $n$ generations of leptons. In
that case we will have $n^{2}+n$ moduli and $n^{2}-n-1$ phases.%
} $12$ moduli ($3$ from $Y$, $3$ from $f$ and $6$ from $g$)
and $5$ phases (all from $g$) and the real and positive parameter
$\mu$ (plus, of course, the rest of the parameters in the scalar
potential). However, we will see later that this convention is not
compatible with the standard parametrization of neutrino masses and
mixings and it will be more convenient to use a slightly different
convention for Yukawa coupling phases: we will also choose $Y$ diagonal
with real and positive elements, then we will choose fermion field
rephasings to remove $3$ phases from the elements of $g_{ab}$, leaving
the elements of $f_{ab}$ complex. Charged scalar rephasing can be
further used to remove the phase of $\mu$ and one of the phases of
$f_{ab}$, for instance we can take $f_{\mu\tau}$ real and positive.
Of course the counting of parameters is the same as before: we will
have $12$ moduli ($3$ from $Y$, $3$ from $f$ and $6$ from $g_{ab}$)
and $5$ phases ($3$ from $g_{ab}$ and $2$ from $f_{ab}$) and
the real and positive parameter $\mu$.

In any of the discussed conventions, $Y$ is directly related to the
masses of charged leptons $m_{a}=Y_{aa}v$, with $v\equiv\left\langle H^{0}\right\rangle =174\,\mathrm{GeV}$,
the VEV of the standard Higgs doublet. Then the physical scalar masses
are\begin{equation}
m_{h}^{2}=m_{h}^{\prime2}+\lambda_{hH}v^{2}\:,\qquad m_{k}^{2}=m_{k}^{\prime2}+\lambda_{kH}v^{2}\,.\end{equation}

\subsection{The neutrino masses}

\label{secmas} 
\begin{figure}
\begin{centering}\includegraphics[width=0.6\columnwidth]{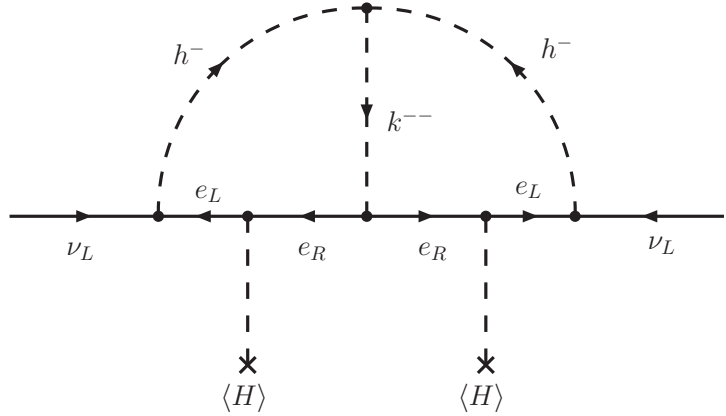}\par\end{centering}

\caption{Diagram contributing to the neutrino Majorana mass at two loops.}

\label{fig:babumass}
\end{figure}

The first contribution to neutrino masses involving
the four relevant couplings appears at two loops \cite{Zee:1985id,Babu:1988ki}
and its Feynman diagram is depicted in fig.~\ref{fig:babumass}.

The calculation of this diagram gives the following mass matrix for
the neutrinos (defined as an effective term in the Lagrangian $\mathcal{L}_{\nu}\equiv-\half\overline{\nu_{L}^{c}}\mathcal{M}_{\nu}\nu_{L}+\mathrm{h.c.}$) 

\begin{equation}
(\mathcal{M}_{\nu})_{ab}=16\mu f_{ac}m_{c}g_{cd}^{*}I_{cd}m_{d}f_{bd}\,,\label{eq:babumas}\end{equation}
 with \begin{equation}
I_{cd}=\int\!\!\frac{\mathrm{d}^{4}k}{(2\pi)^{4}}\int\!\!\frac{\mathrm{d}^{4}q}{(2\pi)^{4}}\,\frac{1}{(k^{2}-m_{c}^{2})}\frac{1}{(k^{2}-m_{h}^{2})}\frac{1}{(q^{2}-m_{d}^{2})}\frac{1}{(q^{2}-m_{h}^{2})}\frac{1}{(k-q)^{2}-m_{k}^{2}}\,.\end{equation}
$I_{cd}$ can be calculated analytically \cite{McDonald:2003zj},
however, since $m_{c},m_{d}$ are the masses of the charged leptons,
necessarily much lighter than the charged scalars, we can neglect
them and obtain a much simpler form\begin{eqnarray}
I_{cd} & \simeq & I=\frac{1}{(16\pi^{2})^{2}}\frac{1}{M^{2}}\frac{\pi^{2}}{3}\tilde{I}(r)\quad,\quad M\equiv\max(m_{h},m_{k})\,,\label{eq:int_mas}\end{eqnarray}
where $\tilde{I}(r)$ is a function of the ratio of the masses of
the scalars $r\equiv m_{k}^{2}/m_{h}^{2}$, \begin{equation}
\tilde{I}(r)=\begin{cases}
1+\frac{3}{\pi^{2}}(\log^{2}r-1) & \text{for }r\gg1\\
1 & \text{for }r\rightarrow0\end{cases}\,,\end{equation}
which is close to one for a wide range of scalar masses. With this
approximation the neutrino mass matrix can be directly written in
terms of the Yukawa coupling matrices, $f$, $g$, and $Y$ \begin{equation}
\mathcal{M}_{\nu}=\frac{v^{2}\mu}{48\pi^{2}M^{2}}\tilde{I}\, f\, Y\, g^{\dagger}Y^{T}f^{T}\,.\label{eq:MnuYukawas}\end{equation}
A very important point is that since $f$ is a $3\times3$ antisymmetric
matrix, $\det f=0$, and therefore $\det\mathcal{M}_{\nu}=0$. Thus,
at least one of the neutrinos is exactly massless at this order%
\footnote{This result does not change if higher orders in charged leptons masses
are taken into account in the loop integral $I_{ab}$. However, one
expects it will change if higher loops are considered.%
}. This is a very important result since it excludes the possibility
of degenerate neutrino masses. 

To estimate the value of the largest possible neutrino mass we can
take $\mu\approx m_{k}\approx m_{h}\equiv M$, then the largest $\nu$
mass will be\begin{equation}
m_{\nu}\approx6.6\times10^{-3}f^{2}g\frac{m_{\tau}^{2}}{M}\,,\end{equation}
which is the typical seesaw formula, suppressed by some additional
couplings and loop factors. Because in this model one of the neutrinos
is massless, the heaviest neutrino mass is fixed by the atmospheric
mass difference, thus $m_{\nu}\approx0.05\,\mathrm{eV}$ and\begin{equation}
f^{2}g\approx150\frac{Mm_{\nu}}{m_{\tau}^{2}}>2\times10^{-7}\,,\end{equation}
since LEP bounds on charged scalar masses are typically $M>100\,\mathrm{GeV}$.
This means that $f$'s and $g$'s cannot be made arbitrarily small
and natural values for them can be $g,f\gtrsim0.01$. Then, for these
relatively large couplings and scalar masses in a range $M\sim100\,\mathrm{GeV-10\, TeV}$
the model will give sizable contributions to low energy processes
like $\mu\rightarrow eee$, $\mu\rightarrow e\gamma$, ..., and scalars
that could be produced and detected at the LHC. 

Alternatively, if we assume that the Yukawa couplings are smaller
than one, $f,g<1$ we could write\begin{equation}
M\approx6.6\times10^{-3}f^{2}g\frac{m_{\tau}^{2}}{m_{\nu}}<4\times10^{5}\,\mathrm{TeV\,,}\end{equation}
which is out of reach of planned experiments. However, constraints
on low energy processes might require smaller couplings which would
lead to much smaller scalar masses. In addition, these estimates are
very rough; for example the relevant couplings may be related to muon
physics, and not to tau physics: our estimate on neutrino masses should
then be reduced by a factor $(m_{\mu}/m_{\tau})^{2}$, which of course
requires much lighter scalars to match the atmospheric neutrino scale.
It is therefore very important to carefully establish the parameters
and validity of the model.

\subsection{Low energy constraints\label{sub:Low-energy-constraints} }

In order to provide neutrino masses compatible with experiment, the
Yukawa couplings of the charged scalars cannot be too small and their
masses cannot be too large. This immediately gives rise to a series
of flavour lepton number violating processes, as for instance $\mu^{-}\rightarrow e^{-}\gamma$
or $\mu^{-}\rightarrow e^{+}e^{-}e^{-}$, with rates which can be,
in some cases, at the verge of the present experimental limits. This
means that we can use these processes to obtain information on the
parameters of the model and perhaps to confirm or to exclude the model
in a close future. In this section we will discuss briefly the relevant
processes and collect the formulas for our conventions of Yukawa couplings:

\begin{itemize}
\item $\ell_{a}^{-}\rightarrow\ell_{b}^{+}\ell_{c}^{-}\ell_{d}^{-}$ : The
interesting observable for these processes is the decay width. We
have (see for instance \cite{Bernabeu:1985na})\begin{equation}
\mathrm{R}(\ell_{a}^{-}\rightarrow\ell_{b}^{+}\ell_{c}^{-}\ell_{d}^{-})\equiv\frac{\Gamma(\ell_{a}^{-}\rightarrow\ell_{b}^{+}\ell_{c}^{-}\ell_{d}^{-})}{\Gamma(\ell_{a}^{-}\rightarrow\ell_{b}^{-}\nu\bar{\nu})}=\frac{1}{2(1+\delta_{cd})}\abs{\frac{g_{ab}g_{cd}^{*}}{G_{F}m_{k}^{2}}}^{2}\,.\end{equation}
In this expression, the term $\delta_{cd}$ takes into account the
fact that we may have two identical particles in the final state.
In the case of $\tau$ decays we have to remember that leptonic channels
are a small fraction of the decays $\mathrm{BR}(\ell_{a}^{-}\rightarrow\ell_{b}^{+}\ell_{c}^{-}\ell_{d}^{-})=\mathrm{R}(\ell_{a}^{-}\rightarrow\ell_{b}^{+}\ell_{c}^{-}\ell_{d}^{-})\mathrm{BR}(\ell_{a}^{-}\rightarrow\ell_{b}^{-}\nu\bar{\nu})$
(with $\mathrm{BR}(\mu^{-}\rightarrow e^{-}\nu\bar{\nu})\approx100\%$,
$\mathrm{BR}(\tau^{-}\rightarrow e^{-}\nu\bar{\nu})\approx17.84\%$~and
$\mathrm{BR}(\tau^{-}\rightarrow\mu^{-}\nu\bar{\nu})\approx17.36\%$).
\item $\mu^{+}e^{-}\longleftrightarrow\mu^{-}e^{+}$: The $k^{++}$ scalar
exchange gives also rise to transitions of the type $\mu^{+}e^{-}\rightarrow\mu^{-}e^{+}$
which are well bounded experimentally. The relevant four-fermion effective
coupling generated by exchange of the scalar $k^{++}$ is (here we
use the conventions for the effective Hamiltonian and the limits of
\cite{Horikawa:1995ae,Willmann:1998gd}) \begin{equation}
G_{M\bar{M}}=-\frac{\sqrt{2}}{8}\frac{g_{ee}g_{\mu\mu}^{*}}{m_{k}^{2}}\,.\end{equation}
We collect the relevant constraints of this type in table \ref{tab:meee}.%
\begin{table}
\begin{tabular}{ccl}
\hline 
Process&
Experiment (90\% CL)&
Bound (90\% CL)\tabularnewline
\hline
$\mu^{-}\rightarrow e^{+}e^{-}e^{-}$&
BR$<1.0\times10^{-12}$&
$|g_{e\mu}g_{ee}^{*}|<2.3\times10^{-5}\,(m_{k}/\mathrm{TeV})^{2}$\tabularnewline
$\tau^{-}\rightarrow e^{+}e^{-}e^{-}$&
BR$<3.6\times10^{-8}$&
$|g_{e\tau}g_{ee}^{*}|<0.010\,(m_{k}/\mathrm{TeV})^{2}$\tabularnewline
$\tau^{-}\rightarrow e^{+}e^{-}\mu^{-}$&
BR$<2.7\times10^{-8}$&
$|g_{e\tau}g_{e\mu}^{*}|<0.006\,(m_{k}/\mathrm{TeV})^{2}$\tabularnewline
$\tau^{-}\rightarrow e^{+}\mu^{-}\mu^{-}$&
BR$<2.3\times10^{-8}$&
$|g_{e\tau}g_{\mu\mu}^{*}|<0.008\,(m_{k}/\mathrm{TeV})^{2}$\tabularnewline
$\tau^{-}\rightarrow\mu^{+}e^{-}e^{-}$&
BR$<2.0\times10^{-8}$&
$|g_{\mu\tau}g_{ee}^{*}|<0.008\,(m_{k}/\mathrm{TeV})^{2}$\tabularnewline
$\tau^{-}\rightarrow\mu^{+}e^{-}\mu^{-}$&
BR$<3.7\times10^{-8}$&
$|g_{\mu\tau}g_{e\mu}^{*}|<0.008\,(m_{k}/\mathrm{TeV})^{2}$\tabularnewline
$\tau^{-}\rightarrow\mu^{+}\mu^{-}\mu^{-}$&
BR$<3.2\times10^{-8}$&
$|g_{\mu\tau}g_{\mu\mu}^{*}|<0.010\,(m_{k}/\mathrm{TeV})^{2}$\tabularnewline
$\mu^{+}e^{-}\rightarrow\mu^{-}e^{+}$&
$G_{M\bar{M}}<0.003G_{F}$&
$|g_{ee}g_{\mu\mu}^{*}|<0.2\,(m_{k}/\mathrm{TeV})^{2}$\tabularnewline
\hline
\end{tabular}

\caption{Constraints from tree-level lepton flavour violating decays.\label{tab:meee}}
\end{table}

\item $\ell_{a}\rightarrow\ell_{b}\nu\bar{\nu}$: These processes receive
additional contributions from the exchange of the singly charged scalar
$h^{+}$ which affect the Fermi muon decay constant but do not modify
the spectrum \cite{Bertolini:1987kz}\begin{equation}
\left(\frac{G_{\mu}}{G_{\mu SM}}\right)^{2}\approx1+\frac{\sqrt{2}}{G_{F}m_{h}^{2}}\abs{f_{e\mu}}^{2}+\frac{1}{2G_{F}^{2}m_{h}^{4}}\left(\abs{f_{e\mu}}^{2}+\abs{f_{e\tau}}^{2}\right)\left(\abs{f_{e\mu}}^{2}+\abs{f_{\mu\tau}}^{2}\right)\,,\end{equation}
where a sum over undetected neutrinos has been performed. The second
term is clearly subdominant if $m_{h}\gg200\,\mathrm{GeV}$, however
we have included it in the numerical analysis and have checked that
we can neglect it in analytical estimates. In this model the charged
scalar only contributes to lepton decays but does not contribute to
hadronic decays, therefore the effective $G_{\beta}$ extracted from
hadronic decays and $G_{\mu}$ are different. However, in the framework
of the SM, the equality of $G_{\beta}$ and $G_{\mu}$ has been tested
with very good accuracy once all radiative corrections have been correctly
included. In the SM, both $|V_{ud}|^{2}+|V_{us}|^{2}+|V_{ub}|^{2}=1$
and $G_{\beta SM}=G_{\mu SM}$ are satisfied. Thus, assuming $G_{\beta SM}=G_{\mu SM}$
one can test the unitarity of the CKM matrix, or, conversely, assuming
the unitarity of the CKM matrix one can test the universality of couplings
in hadronic and leptonic decays. In the model we are considering this
is not true anymore; the CKM matrix is still unitary but, as explained,
$G_{\beta}=G_{\beta SM}=G_{\mu SM}\not=G_{\mu}$. Since the extraction
of the experimental values of $|V_{ij}^{exp}|^{2}$ assumes the SM,
we will have \begin{equation}
V_{ij}^{exp}=\frac{G_{\beta}}{G_{\mu}}V_{ij}\,,\end{equation}
where $V_{ij}$ are the truly unitary CKM matrix elements in the model.
Therefore \begin{equation}
|V_{ud}^{exp}|^{2}+|V_{us}^{exp}|^{2}+|V_{ub}^{exp}|^{2}=\frac{G_{\beta}^{2}}{G_{\mu}^{2}}=\frac{G_{\mu SM}^{2}}{G_{\mu}^{2}}\approx1-\frac{\sqrt{2}}{G_{F}m_{h}^{2}}\abs{f_{e\mu}}^{2}\,,\end{equation}
and, since $|V_{ud}^{exp}|^{2}+|V_{us}^{exp}|^{2}+|V_{ub}^{exp}|^{2}=0.9992\pm0.0011$
\cite{Yao:2006px} is very close to $1$, we will obtain a strong
limit on $|f_{e\mu}|^{2}/m_{h}^{2}$.\\
On the other hand, the charged scalar contribution will also modify
the Fermi coupling extracted from $\tau$ decays in the different
leptonic channels. After subtracting the different factors from phase
space and radiative corrections this is usually expressed in terms
of ratios of effective {}``gauge couplings'' $g_{a}^{exp}$ for
the different leptons which in the SM are all equal (see for instance
\cite{Pich:2006nt}). Thus, comparing tau decays to muons and tau
decays to electrons we have (since in the SM $G_{a\rightarrow b}\propto g_{a}g_{b}$)
\begin{equation}
\left(\frac{g_{\mu}^{exp}}{g_{e}^{exp}}\right)^{2}=\left(\frac{G_{\tau\rightarrow\mu}}{G_{\tau\rightarrow e}}\right)^{2}\approx1+\frac{\sqrt{2}}{G_{F}m_{h}^{2}}\left(\abs{f_{\mu\tau}}^{2}-\abs{f_{e\tau}}^{2}\right)\,.\end{equation}
Similarly\begin{equation}
\left(\frac{g_{\tau}^{exp}}{g_{\mu}^{exp}}\right)^{2}=\left(\frac{G_{\tau\rightarrow e}}{G_{\mu\rightarrow e}}\right)^{2}\approx1+\frac{\sqrt{2}}{G_{F}m_{h}^{2}}\left(\abs{f_{e\tau}}^{2}-\abs{f_{e\mu}}^{2}\right)\,,\end{equation}
\begin{equation}
\left(\frac{g_{\tau}^{exp}}{g_{e}^{exp}}\right)^{2}=\left(\frac{G_{\tau\rightarrow\mu}}{G_{\mu\rightarrow e}}\right)^{2}\approx1+\frac{\sqrt{2}}{G_{F}m_{h}^{2}}\left(\abs{f_{\mu\tau}}^{2}-\abs{f_{e\mu}}^{2}\right)\,.\end{equation}

\end{itemize}
Universality constraints are summarized in table \ref{tab:universality}%
\begin{table}
\begin{tabular}{ccc}
\hline 
SM Test&
Experiment&
Bound (90\%CL)\tabularnewline
\hline
lept./hadr. univ.&
$\sum_{q=d,s,b}|V_{uq}^{exp}|^{2}=0.9992\pm0.0011$&
$|f_{e\mu}|^{2}<0.015\,(m_{h}/\mathrm{TeV})^{2}$\tabularnewline
$\mu/e$ universality&
$g_{\mu}^{exp}/g_{e}^{exp}=1.0001\pm0.0020$&
$\abs{|f_{\mu\tau}|^{2}-|f_{e\tau}|^{2}}<0.05\,(m_{h}/\mathrm{TeV})^{2}$\tabularnewline
$\tau/\mu$ universality&
$g_{\tau}^{exp}/g_{\mu}^{exp}=1.0004\pm0.0022$&
$\abs{|f_{e\tau}|^{2}-|f_{e\mu}|^{2}}<0.06\,(m_{h}/\mathrm{TeV})^{2}$\tabularnewline
$\tau/e$ universality&
$g_{\tau}^{exp}/g_{e}^{exp}=1.0004\pm0.0023$&
$\abs{|f_{\mu\tau}|^{2}-|f_{e\mu}|^{2}}<0.06\,(m_{h}/\mathrm{TeV})^{2}$\tabularnewline
\hline
\end{tabular}

\caption{Constraints from universality of charged currents.\label{tab:universality}}
\end{table}
, where measured values are translated into $90$\%~CL limits.

\begin{itemize}
\item $\ell_{a}^{-}\rightarrow\ell_{b}^{-}\gamma$ : In the case of transition
amplitudes $a\not=b$ the interesting observable is the decay rate.
We find (for calculations including singly and doubly charged scalars
see for instance \cite{Bilenky:1987ty,Bertolini:1987kz,Pich:1985uv})
\begin{equation}
\mathrm{R}(\ell_{a}^{-}\rightarrow\ell_{b}^{-}\gamma)\equiv\frac{\Gamma(\ell_{a}^{-}\rightarrow\ell_{b}^{-}\gamma)}{\Gamma(\ell_{a}^{-}\rightarrow\ell_{b}^{-}\nu\bar{\nu})}\approx\frac{\alpha}{48\pi}\left(\abs{\frac{(f^{\dagger}f)_{ab}}{G_{F}m_{h}^{2}}}^{2}+16\abs{\frac{(g^{\dagger}g)_{ab}}{G_{F}m_{k}^{2}}}^{2}\right)\,.\end{equation}
The factor $16$ in front of the doubly charged contribution does
not usually appear in the literature \cite{Babu:2002uu} and deserves
a comment: the Feynman rule for the $ke_{a}e_{b}$ vertex contains
a factor $2$ when $a\not=b$ because there are two identical terms
in the Lagrangian, but also the vertex with $a=b$ contains a factor
$2$ because there are two identical Wick contractions in this kind
of vertices. This factor of $2$ for identical particles was missed
in \cite{Babu:2002uu} which led the authors do define new coupling
constants with different factors of $2$ for diagonal and non-diagonal
terms%
\footnote{One can also see that those results cannot be right because physical
amplitudes should transform correctly under the flavour redefinitions
of couplings in eq.~\eqref{eq:redefinitions}.%
}. It is also important to remark that the singly and doubly charged
scalar contributions do not interfere because they couple to fermions
with different chirality. Again we have to remember that $\mathrm{BR}(\ell_{a}^{-}\rightarrow\ell_{b}^{-}\gamma)=\mathrm{R}(\ell_{a}^{-}\rightarrow\ell_{b}^{-}\gamma)\mathrm{BR}(\ell_{a}^{-}\rightarrow\ell_{b}^{-}\nu\bar{\nu})$. 
\item $\mu-e$ conversion in nuclei: The new scalars of the model do not
couple to quarks and, therefore, do not generate a four-fermion operator
that could contribute at tree level to $\mu-e$ conversion. However,
radiative corrections, in particular those related with the $\mu-e-\gamma$
vertex, will contribute to the process. It is also clear that those
corrections are tightly related to the $\mu\rightarrow e\gamma$ decay
discussed above but are not identical because the photon in $\mu-e$
conversion is not on the mass shell. In fact in ref.~\cite{Raidal:1997hq}
if was shown that in models with doubly charged scalars there is a
logarithmic enhancement, $\log(q^{2}/m_{k})$, of the $\mu-e$ conversion
amplitude with respect to the $\mu\rightarrow e\gamma$ amplitude.
At present the best limits come from $\mu-e$ conversion on $Ti$,
$\sigma(\mu^{-}Ti\rightarrow e^{-}Ti)/\sigma(e^{-}Ti\rightarrow\mathrm{capture})<4.3\times10^{-12}$
\cite{Yao:2006px} which, when translated into limits on the couplings,
are slightly worse than present $\mu\rightarrow e\gamma$ constraints,
but one has to keep in mind that if $\mu\rightarrow e\gamma$ is relevant
and if $\mu-e$ conversion limits are improved in the future it will
also be relevant.
\item $a=(g-2)/2$: For diagonal transitions the model gives additional
contributions to the anomalous magnetic moments of the leptons which
are very well measured in the case of the electron and the muon. We
find that the additional contribution to the $a_{a}$ of lepton $\ell_{a}$,
in this model, is\begin{equation}
\delta a_{a}\equiv a_{a}^{exp}-a_{a}^{SM}=-\frac{m_{a}^{2}}{24\pi^{2}}\left(\frac{(f^{\dagger}f)_{aa}}{m_{h}^{2}}+4\frac{(g^{\dagger}g)_{aa}}{m_{k}^{2}}\right)\,.\end{equation}
It is important to remark that we always find a negative contribution
(this is in agreement with old and well tested calculations \cite{Moore:1984eg,Leveille:1977rc}).
In the case of the muon $a_{\mu}$, recent analyses of experimental
data and theoretical calculations in the Standard Model suggest that
the experimental measurement is slightly larger than the SM prediction
(for a review see \cite{Passera:2004bj}). Several authors have tried
to explain this $1\sigma$ to $3\sigma$ effect in different extensions
of the SM. In particular, in \cite{Dicus:2001ph} the charged scalars
of the Zee model were used to increase the $a_{\mu}$ of the SM. We
find this is not possible and instead we will use the $g-2$ of the
muon (and also of the electron) to constrain the parameters of the
model. The relevant constraints coming from flavour lepton number
changing photon interactions are summarized in table \ref{tab:meg}%
\begin{table}
\begin{tabular}{cl}
\hline 
Experiment&
~~~~~Bound (90\%CL)\tabularnewline
\hline
$\delta a_{e}=(12\pm10)\times10^{-12}$&
$r\left(|f_{e\mu}|^{2}+|f_{e\tau}|^{2}\right)+4\left(|g_{ee}|^{2}+|g_{e\mu}|^{2}+|g_{e\tau}|^{2}\right)<5.5\times10^{3}\,(m_{k}/\mathrm{TeV})^{2}$\tabularnewline
$\delta a_{\mu}=(21\pm10)\times10^{-10}$&
$r\left(|f_{e\mu}|^{2}+|f_{\mu\tau}|^{2}\right)+4\left(|g_{e\mu}|^{2}+|g_{\mu\mu}|^{2}+|g_{\mu\tau}|^{2}\right)<7.9\,(m_{k}/\mathrm{TeV})^{2}$\tabularnewline
$BR(\mu\rightarrow e\gamma)<1.2\times10^{-11}$&
$r^{2}|f_{e\tau}^{*}f_{\mu\tau}|^{2}+16|g_{ee}^{*}g_{e\mu}+g_{e\mu}^{*}g_{\mu\mu}+g_{e\tau}^{*}g_{\mu\tau}|^{2}<3.4\times10^{-5}\,(m_{k}/\mathrm{TeV})^{4}$\tabularnewline
$BR(\tau\rightarrow e\gamma)<1.1\times10^{-7}$&
$r^{2}|f_{e\mu}^{*}f_{\mu\tau}|^{2}+16|g_{ee}^{*}g_{e\tau}+g_{e\mu}^{*}g_{\mu\tau}+g_{e\tau}^{*}g_{\tau\tau}|^{2}<1.7\,(m_{k}/\mathrm{TeV})^{4}$\tabularnewline
$BR(\tau\rightarrow\mu\gamma)<4.5\times10^{-8}$&
$r^{2}|f_{e\mu}^{*}f_{e\tau}|^{2}+16|g_{e\mu}^{*}g_{e\tau}+g_{\mu\mu}^{*}g_{\mu\tau}+g_{\mu\tau}^{*}g_{\tau\tau}|^{2}<0.7\,(m_{k}/\mathrm{TeV})^{4}$\tabularnewline
\hline
\end{tabular}

\caption{Constraints from lepton number violating photon interactions.\label{tab:meg}}
\end{table}
. Experimental limits on branching ratios are already provided in
the literature \cite{Yao:2006px} at $90$\%~CL. Results for $g-2$
are usually given as measurements on $\delta a_{a}$. We will use
\cite{Kinoshita:2006kh} $\delta a_{e}=(12\pm10)\times10^{-12}$ and
$\delta a_{\mu}=(21\pm10)\times10^{-10}$. Notice that in both cases
the central value is positive, while the model gives a negative contribution.
To place $90$\%~CL bounds in this situation we use the Feldman and
Cousins prescription \cite{Feldman:1997qc} which, for the values
above, gives $|\delta a_{e}|<6.1\times10^{-12}$ and $|\delta a_{\mu}|<3.8\times10^{-10}$.
\end{itemize}
Since lepton number is not conserved, another interesting low energy
process that could arise in the model is neutrinoless double beta
decay ($0\nu2\beta$). In this model, the singly and doubly charged
scalars do not couple to hadrons, this means that the ($0\nu2\beta$)
rate is dominated by the Majorana neutrino mass exchange and it is
proportional to the $|(\mathcal{M}_{\nu})_{ee}|^{2}$ matrix element,
therefore it will be addressed in section~\ref{sec:Analysis} when
we discuss in detail the neutrino mass matrix constraints.

\subsection{Perturbativity constraints \label{sub:Perturbativity-constraints}}

Beside requiring that the model produce acceptable (a) neutrino masses
and mixings (b) low energy predictions, we also have to address theoretical
questions related to the validity of the predictions and the consistency
of the model. Indeed to be able to perform any calculation in this
model we have to assume that perturbation theory can be used. This
imposes strong constraints on the relevant couplings of the model.
The Yukawa couplings of the new scalars receive loop corrections like

\begin{equation}
\delta f\sim\frac{f^{3}}{(4\pi)^{2}}\,,\qquad\delta g\sim\frac{g^{3}}{(4\pi)^{2}}\,.\end{equation}
If the corrections are going to be much smaller than the couplings,
the couplings must satisfy $f,g\ll4\pi$. Similarly, the trilinear
coupling among charged scalars proportional to the parameter $\mu$
induces loop corrections to the charged scalar masses like 

\begin{equation}
\delta m_{k}^{2},\delta m_{h}^{2}\sim\frac{\mu^{2}}{(4\pi)^{2}}\,.\end{equation}
Requiring that the corrections are much smaller than the masses implies
$\mu\ll4\pi m_{h},4\pi m_{k}$. Since it is difficult to establish
the exact values of the couplings for which the perturbativity of
the theory breaks down, we will encode this type of constraints in
the single parameter $\kappa$ and will require \begin{equation}
|f_{ab}|<\kappa\,,\qquad|g_{ab}|<\kappa\,,\qquad\mu<\kappa\min(m_{h},m_{k})\,.\label{eq:perturbativity}\end{equation}
For the purposes of illustration we will take $\kappa=1$ or $\kappa=5$,
the value $\kappa=5$ being rather conservative (for instance, the
strong coupling constant, $g_{s}$, is considered to become non-perturbative
at scales of about $1\,\mathrm{GeV}$, at those scales $\alpha_{s}(1\,\mathrm{GeV)\sim0.5}$
and $g_{s}(1\,\mathrm{GeV)\sim2.5}$). 

The parameter $\mu$ is important in the generation of neutrino masses,
therefore the constraint $\mu<\kappa\min(m_{h},m_{k})$ is important.
On the other hand, as seen in tables~\ref{tab:meee}, \ref{tab:universality},
\ref{tab:meg}, if the scalar masses are relatively light (around
$1$~TeV or less), low energy processes already provide interesting
limits on the charged scalar Yukawa couplings. However, if the charged
scalars are heavier, the experimental limits on the new Yukawa couplings
are so mild that may allow Yukawa couplings large enough to compromise
the perturbative validity of the theory. Then, the perturbativity
constraints we just discussed will become relevant.

\section{The model at the LHC\label{sec:LHC}}

Extra scalar degrees of freedom arise in many scenarios extending
the weak interactions beyond the SM. In our case, the scalar sector
is enlarged by the addition of two charged scalars: $h$ and $k$,
which could be produced at the LHC if their masses are low enough.
In particular, as we will see below, the LHC will be very well suited
for searching the doubly charged scalar, $k$. Studies for searching
doubly-charged scalars at future colliders have been directed in the
past \cite{Gunion:1989in,Huitu:1996su,Gunion:1996pq,Akeroyd:2005gt,Azuelos:2005uc}
. In general this scalar is taken to be one component of a weak triplet.
Such triplets are well motivated on theoretical grounds, specially
when considering left-right symmetric models. Our work differs essentially
in the gauge charges of the scalars. Both, $h$ and $k$, are charged
weak singlets that do not acquire a VEV. This makes the phenomenology
different. More model independent studies have been also considered
in the literature \cite{Dion:1998pw,Cuypers:1996ia}.

Concerning experimental bounds, LEP searched for these scalars. Their
pair production ($e^{+}e^{-}\rightarrow\gamma^{\ast}Z^{\ast}\rightarrow k^{++}k^{--}$)
implies the bound $m_{k}>100~\mbox{GeV}$ \cite{Abdallah:2002qj,Abbiendi:2001cr,Achard:2003mv}.
Single production via $e^{+}e^{-}\rightarrow kee$ as well as contributions
to Bhabha scattering have been also studied by LEP \cite{Achard:2003mv,Abbiendi:2003pr},
but in these cases the bounds depend on the (unknown) values of the
Yukawa couplings. Tevatron has also been used to set bounds on this
kind of scalars \cite{Abazov:2004au,Acosta:2004uj,Acosta:2005np}.
Depending on the details of the model (couplings, decay channels,...)
the mass is again found to be roughly above $100~\mbox{GeV}$.

\subsection{Collider phenomenology}

\label{sec:colpheno}

\subsubsection{Production}

\begin{figure}
\begin{centering}\includegraphics[width=0.4\columnwidth]{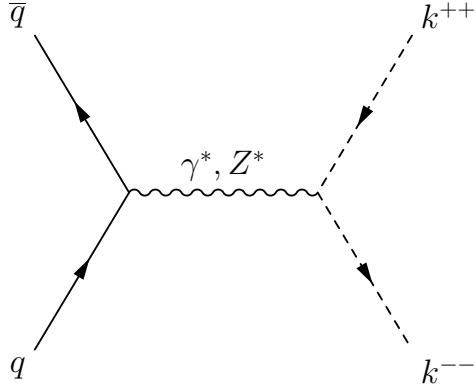}\par\end{centering}

\caption{Pair production of $k$\label{fig:pp}}
\end{figure}

The extra scalars can be pair produced via a Drell-Yan process, fig.~\ref{fig:pp}.
Although this production mechanism presents the drawback of having
a potentially high threshold due to the creation of two scalars, it
has the important advantage of being proportional to their gauge charges
as well as depending only on one unknown parameter: the mass of the
scalar. The partonic cross section at LO reads \begin{equation}
\sigma=\frac{\pi\alpha^{2}Q^{2}\beta^{3}}{6}\left[\frac{2Q_{q}^{2}}{\hat{s}}-\frac{2(g_{L}+g_{R})Q_{q}}{c_{w}^{2}}\frac{\hat{s}-M_{Z}^{2}}{(\hat{s}-M_{Z}^{2})^{2}+\Gamma_{Z}^{2}M_{Z}^{2}}+\frac{(g_{L}^{2}+g_{R}^{2})}{c_{w}^{4}}\frac{\hat{s}}{(\hat{s}-M_{Z}^{2})^{2}+\Gamma_{Z}^{2}M_{Z}^{2}}\right],\label{eq:pxsec}\end{equation}
 where $\hat{s}$ is the energy squared in the center of mass frame
(CM) of the quarks, $Q$ stands for electric charges, $g_{L}$ and
$g_{R}$ are given for the quarks by $g_{L}=T_{3}-s_{w}^{2}Q_{q}$
and $g_{R}=-s_{w}^{2}Q_{q}$ and $\beta$ is the velocity of the produced
scalars in this frame $\beta=\sqrt{1-4m^{2}/\hat{s}}$.

\begin{figure}
\begin{centering}\includegraphics[width=0.7\columnwidth]{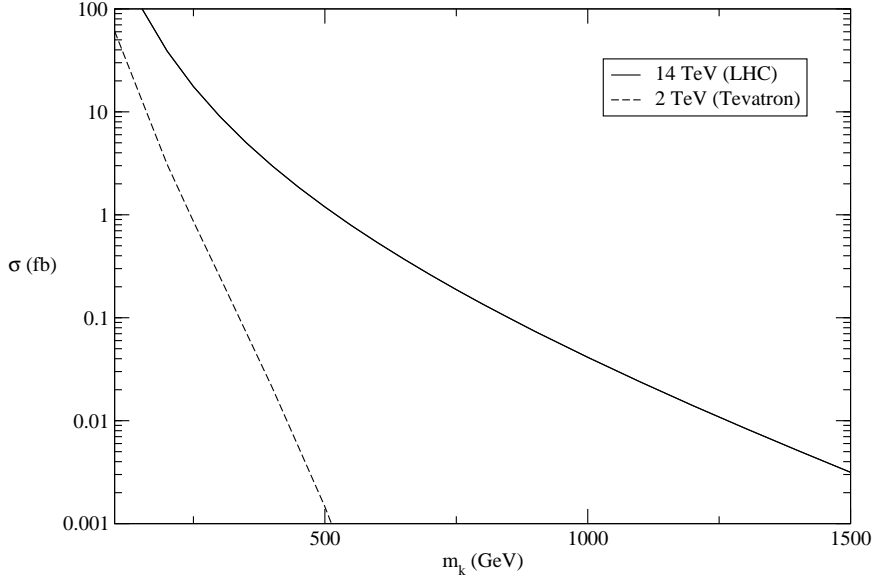}\par\end{centering}

\caption{Pair production cross section for $k$. We have used CompHEP (CTEQ6.1L)
to obtain the LO and applied a K-factor of $1.25$ for the LHC and
$1.3$ for Tevatron.\label{fig:ppxsecs}}
\end{figure}

Equation~\eqref{eq:pxsec} shows that pair production is four times
more efficient for $k$ than for $h$ due to their charges (assuming
equal masses), which translates into a better discovery potential
for $k$. The $k$ pair production cross section, $\sigma_{kk}$,
at NLO for the LHC and Tevatron is displayed in fig.~\ref{fig:ppxsecs}.
To compute it, we have used CompHEP~\cite{Pukhov:1999gg} with CTEQ6.1L
libraries~\cite{Stump:2003yu} to find the LO cross section and afterwards
we have included a K-factor of $1.25$ for the LHC and $1.3$ for
Tevatron to take into account NLO corrections, see \cite{Muhlleitner:2003me}.

\begin{figure}
\begin{centering}\includegraphics[width=0.45\columnwidth]{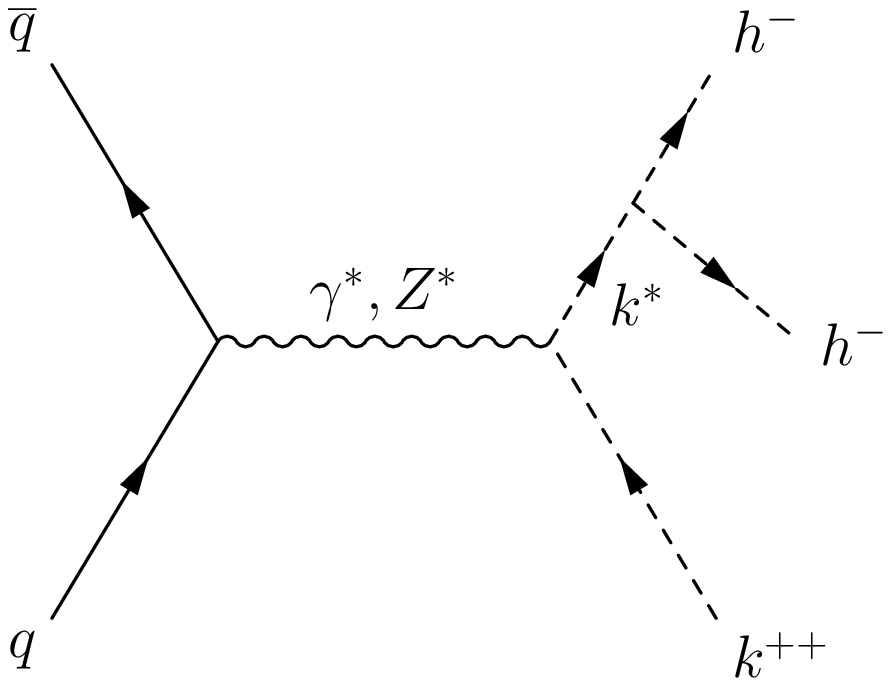} ~~~~~\includegraphics[width=0.45\columnwidth]{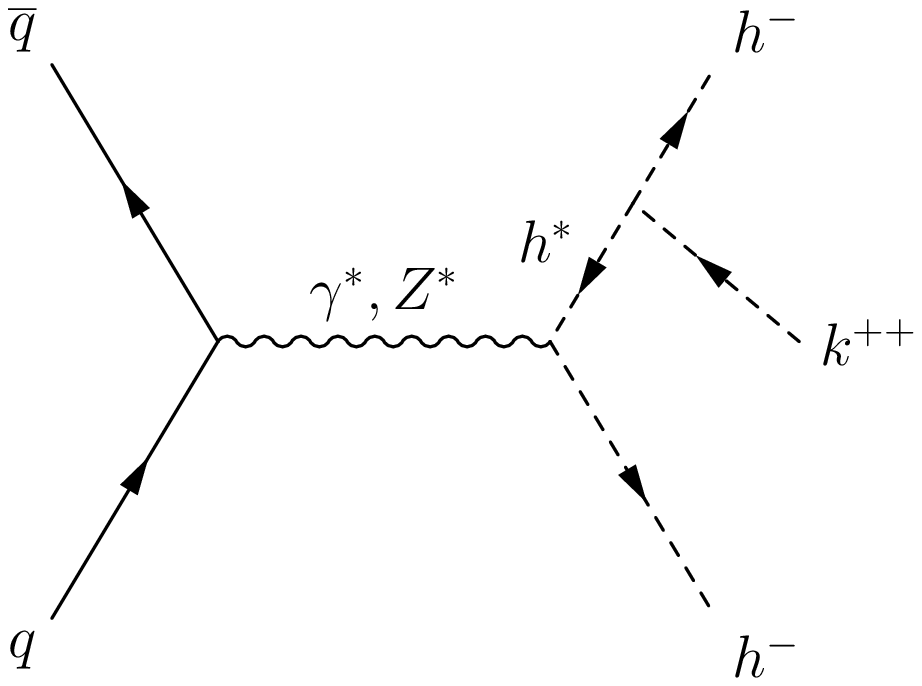}\par\end{centering}

\caption{Single production diagrams.\label{fig:singprod}}
\end{figure}

Single production might be also interesting when double production
is not possible. Single production can proceed with a $k$ accompanied
by two singly charged scalars, fig.~\ref{fig:singprod}, or by two
charged leptons replacing the scalar $h$'s. If the $k$ is accompanied
by two charged leptons the amplitudes are proportional to the Yukawa
couplings, whose exact values we ignore and might be small. %

It is important to note that the cross section will be dominated by
the virtual particles in the propagators if they could be on-shell.
In the case of $k$ being produced with two $h$, the single production
will be dominated by the first diagram if $\hat{s}>2m_{k}$, because
in this case $k^{\ast}$ can be created on-shell. One might argue
that the energy in the center of mass frame of the colliding quarks
is not fixed, instead it is a fraction of the total energy in the
center of mass frame of the colliding protons, $s$. However, the
cross section involves an integration over the possible values of
$\hat{s}$. If $s$ is large enough to create two $k$'s, the integration
will be dominated by the real production of two $k$'s, thus reducing
the single production to pair production. Specifically, $\sigma(k^{++}h^{-}h^{-})\approx\sigma_{kk}Br(k\rightarrow hh)$.
The same reasoning is valid in the case of single production with
leptons. We have performed calculations using CompHEP to check this
point. Therefore, single production is only important when the available
energy, $s$, is not sufficient to create a pair of $k$.

A possible third production mechanism is via the couplings with the
Higgs doublet, $H$. There is little information concerning these
couplings because their contribution to low energy phenomenology is
expected to be negligible in front of the Yukawas $f_{ab}$ and $g_{ab}$.
This is so because the former enter at two loops and the later at
tree level. In any case, the amplitudes of these processes are expected
to be small because the Higgs couplings to quarks are proportional
to their masses.

In summary, we find that, from the point of view of production, the
best suited channel for discovery studies is pair production, being
four times more efficient for $k$ than for $h$.

\subsubsection{Decay}

We assume for the moment that the scalars are not long lived, i.e.
they decay before reaching the detector. Different decay channels
present different experimental sensitivities depending on the final
products. In particular, detectors are much less sensitive to those
channels containing neutrinos and/or taus in the final states, since
neutrinos (including those coming from the decay of the taus) will
escape undetected. This makes necessary to compute the branching ratios.

The $k$ scalar can always decay to two leptons of the same sign,
since $m_{k}>100\,\mbox{GeV}$. The width reads \begin{equation}
\Gamma(k\rightarrow\ell_{a}\ell_{b})=\frac{|g_{ab}|^{2}}{4\pi(1+\delta_{ab})}m_{k}.\end{equation}
 It is worth to stress that $k$ is the only particle in this model
that can decay to two like-sign leptons, which will be crucial to
detect it.

If $m_{k}>2m_{h}$, then $k$ can also decay to a pair of $h$\begin{equation}
\Gamma(k\rightarrow hh)=\frac{1}{8\pi}\left[\frac{\mu}{m_{k}}\right]^{2}m_{k}\;\sqrt{1-\frac{4m_{h}^{2}}{m_{k}^{2}}}.\end{equation}

On the other hand, the decay channels of $h$ reduce to those with
one lepton and one neutrino in the final state: \begin{equation}
\Gamma(h\rightarrow\ell_{a}\nu_{b})=\frac{|f_{ab}|^{2}}{4\pi}m_{h}.\end{equation}
 Since these channels involve always one neutrino it is clear that
detecting $h$ will be much more complicated than detecting $k$ even
if their production rates were similar.

Finally we will check that the doubly charged scalar $k^{++}$ cannot
be long-lived in this model%
\footnote{By {}``long-lived'' we understand that the scalar can travel a distance
of the order $c\tau>3\, m$~\cite{Acosta:2005np}.%
} once low energy phenomenology and neutrino data are taken into account.
Indeed the $k$ decay width can be written as \begin{equation}
\Gamma=\frac{m_{k}}{8\pi}\,\,\left(|g_{\mu\mu}|^{2}+\cdots\right).\end{equation}
where the $g_{\mu\mu}^{2}$ term takes into account the decay into
muons and the dots represent all other possible couplings. Then, the
long-lived condition translates into $|g_{\mu\mu}|<10^{-8}$ which
cannot be fulfilled when the limits obtained in tabs.~\ref{tab:Normal-Hierarchy-analytical}-\ref{tab:Inverted-Hierarchy-analytical}
are used.

\subsubsection{Detection at colliders}

As we have seen the discovery potential of the LHC for $k$ is more
promising than for $h$. On one side because the production cross
section of the former is enhanced with respect to the later and, on
the other side, because the experimental sensitivity to the decay
channels of $h$ is smaller. Thus, in the following we will focus
on $k$.

After a pair of $k$'s is created in the collider, they can decay
into a number of final states. The most interesting for us contains
four like-sign leptons. From now on we will refer to this channel
as $4lep$. The rest of the possible final states always contain $h$
or $\tau$. These channels are quite difficult to deal with experimentally
because $\tau$ and $h$ will decay to neutrinos. In contrast, detection
of electrons and muons is quite efficient. In addition, the decay
of a $k$ to a pair of like-sign leptons ($e^{\pm}$ or $\mu^{\pm}$)
with high invariant masses constitutes a clear and distinct signature.
This channel has a negligible background coming from SM processes,
making it very appropriate for $k$ discovery studies.

In order to model the efficiencies and acceptances of the detectors
at the LHC we use the criterion that 10 events of $k$ pair production
with subsequent decay to $4lep$ lead to discovery of $k$. We expect
that at least two of such events will be properly detected/identified
providing us with four pairs of like-sign leptons which invariant
mass will give us the first estimate of $m_{k}$. This criterion is
taken from \cite{Gunion:1996pq} where the authors perform a study
of the discovery reach at Tevatron for doubly-charged bosons decaying
to like-sign leptons in a similar model and a similar rule can be
extracted from \cite{Azuelos:2005uc} where the authors focus on the
ATLAS detector at the LHC. A more detailed study of the forthcoming
detectors acceptances and efficiencies at the LHC is desirable.

\begin{figure}
\begin{centering}\includegraphics[width=0.6\columnwidth,angle=-90]{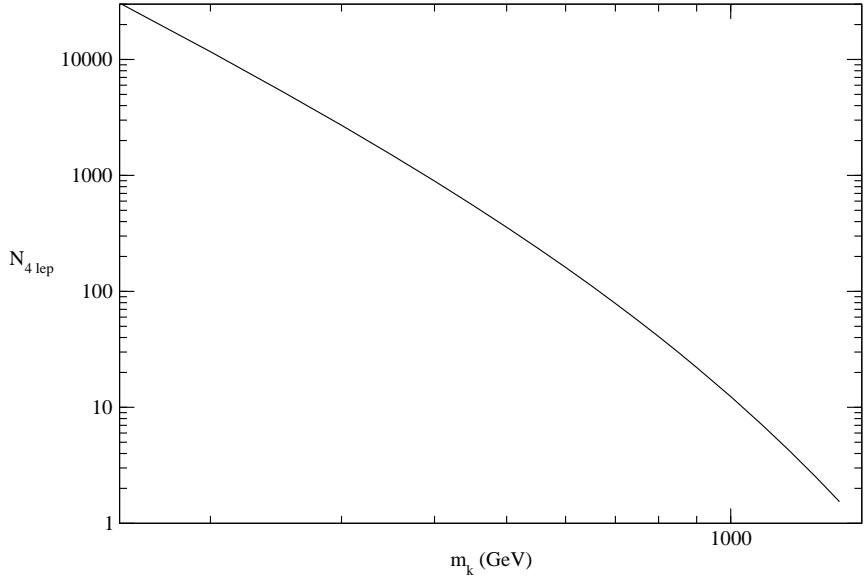}\par\end{centering}

\caption{Number of events at the LHC in the $4lep$ channel for a luminosity
$\mathcal{L}=300~\mbox{fb}^{-1}$ and $\sqrt{s}=14~\mbox{TeV}$ assuming
that all produced $k$ pairs decay in this particular channel.\label{fig:numevents}}
\end{figure}

To estimate the maximum reach in terms of $m_{k}$ at the LHC we take
the most optimistic scenario in which all the $k$ pairs decay to
$4lep$. The number of events in this channel is shown in fig.~\ref{fig:numevents}
for the optimistic luminosity%
\footnote{The LHC luminosity is expected to be about $100\,\mathrm{fb^{-1}}/\mathrm{year}.$%
} $300\,\mbox{fb}^{-1}$ and CM energy $14\,\mbox{TeV}$. From this
plot one concludes that the LHC will be able to probe masses up to
$1\:\mbox{TeV}$ approximately.

In general, the signal in the $4lep$ channel will be smaller than
the one shown in fig.~\ref{fig:numevents} due to the presence of
the other decay channels. This signal draining will be controlled
by the branching ratio $BR_{4lep}$, which can be expressed in terms
of the couplings as \begin{equation}
BR_{4lep}=\frac{|g_{ee}|^{2}+|g_{\mu\mu}|^{2}+2|g_{e\mu}|^{2}}{|g_{ee}|^{2}+|g_{\mu\mu}|^{2}+2|g_{e\mu}|^{2}+|g_{hh}|^{2}+2|g_{e\tau}|^{2}+2|g_{\mu\tau}|^{2}+|g_{\tau\tau}|^{2}}\label{eq:br4l}\end{equation}
 where we have defined the effective coupling of the doubly charged
scalar to singly charged scalars, $g_{hh}$, as \begin{equation}
g_{hh}=\left[\frac{\mu}{m_{k}}\right]\;\left(1-\frac{4m_{h}^{2}}{m_{k}^{2}}\right)^{1/4}.\end{equation}

\section{Analysis of the parameters of the model \label{sec:Analysis}}

As discussed in section \ref{sec:The-Babu-model} the Yukawa couplings
of the model can be written in terms of $12$ moduli and $5$ phases.
Other parameters relevant for the model are the masses of the charged
scalars, $m_{h}$ and $m_{k}$, and the coupling $\mu$. Among the
$12$ moduli, $3$ correspond to the $3$ charged lepton masses, which
are known. Thus, we have $17$ additional relevant parameters with
respect to the plain SM with massless neutrinos ($9$ moduli and $5$
phases from the new Yukawa couplings, the $3$ scalar parameters,
$m_{h}$ and $m_{k}$ and the coupling $\mu$). The remaining parameters
in the scalar potential are of no interest for our purposes. The neutrino
mass matrix is rather well known. In our case it contains $2$ neutrino
masses, $3$ real mixing angles and $2$ phases ($1$ CKM-type phase
and $1$ Majorana phase). Thus, there will still remain $4$ moduli
and $3$ additional phases in the Yukawa couplings (plus $m_{h}$,
$m_{k}$ and $\mu$). On most of these Yukawa couplings we have some
information from section \ref{sub:Low-energy-constraints} as long
as the masses of the charged scalars are not much heavier than 1~TeV.
Notice that this is the interesting range for scalar masses if they
are going to be produced at the LHC. In addition, there are also indirect
arguments that suggest that the scalar masses should be relatively
light (below $1$~TeV) if one likes to avoid strong hierarchy problems%
\footnote{Alternatively one could enlarge the model by supersymmetrizing it.%
}, since the charged scalar masses will contribute, at one loop, to
the mass of the SM Higgs boson. However, the couplings of the SM Higgs
boson to the new scalars are unknown and could be small. Thus, although
the natural range of the masses of the new scalars is about few TeV
or less, they can also be larger. Then, in what follows, we will allow
the masses of the charged scalar to vary between the LEP lower bound
$\sim100\,\mathrm{GeV}$ and infinity. We will immediately see, however,
that present information already constrains the charged scalar masses
to be below $\sim10^{5}\,\mathrm{TeV}$. The couplings $g_{ab}$,
$f_{ab}$ and $\mu$ must in addition satisfy the perturbativity constraints
discussed in section~\ref{sub:Perturbativity-constraints}.

From the previous discussion it is clear that even though we have
$17$ additional parameters we also have a lot of information on them
both from neutrino oscillations and from low energy processes. However,
the correlations among the different observables due to their dependence
on the same set of parameters can be difficult to disentangle. Under
those circumstances an adequate approach to the analysis of the Zee-Babu
model should involve both an analytic understanding and a systematic
numerical inspection of the parameter space to clarify the ranges
allowed by the available experimental results for as many parameters
and observables as possible. 

We first exploit the information we have on neutrino oscillation experiments:
the knowledge of two squared mass differences and the mixing angles. 

With antisymmetric $f_{ab}$'s, at two loops, the mass matrix determinant
is equal to zero, and thus one eigenvalue is zero. Two mass differences
are then sufficient to fix the masses both in the normal hierarchy
and in the inverted hierarchy cases (a degenerate spectrum cannot
arise in this model). Except for the Dirac phase $\delta$ and the
Majorana phase $\phi$, we can almost reconstruct experimentally the
neutrino mass matrix by using the information we have on the mixing
angles and the masses. Without loss of generality we can write the
neutrino Majorana mass matrix as\begin{equation}
\mathcal{M}_{\nu}=UD_{\nu}U^{T}\,,\label{eq:MnuUs}\end{equation}
with $U$ the standard PMNS matrix \begin{equation}
U=\left(\begin{array}{ccc}
1 & 0 & 0\\
0 & c_{23} & s_{23}\\
0 & -s_{23} & c_{23}\end{array}\right)\left(\begin{array}{ccc}
c_{13} & 0 & s_{13}e^{-i\delta}\\
0 & 1 & 0\\
-s_{13}e^{i\delta} & 0 & c_{13}\end{array}\right)\left(\begin{array}{ccc}
c_{12} & s_{12} & 0\\
-s_{12} & c_{12} & 0\\
0 & 0 & 1\end{array}\right)\,,\label{eq:U}\end{equation}
while $D_{\nu}$ is the diagonal matrix of masses (including the only
Majorana phase). Notice that writing the mass matrix in this form
already implies some phase convention.

Since one of the $\nu$ masses of the model is zero we only have two
possibilities%
\footnote{Here we follow the conventions and results of ref.~\cite{Schwetz:2006dh}
adapted to our case.%
}:

Normal hierarchy (NH)\begin{equation}
D_{\nu}^{NH}=\left(\begin{array}{ccc}
0 & 0 & 0\\
0 & m_{2}e^{i\phi} & 0\\
0 & 0 & m_{3}\end{array}\right)\;,m_{3}\gg m_{2}\quad,\begin{array}{c}
\Delta_{S}=m_{2}^{2}\\
\Delta_{A}=m_{3}^{2}\end{array}\,.\label{eq:DiagonalNH}\end{equation}
Inverted hierarchy (IH)\begin{equation}
D_{\nu}^{IH}=\left(\begin{array}{ccc}
m_{1} & 0 & 0\\
0 & m_{2}e^{i\phi} & 0\\
0 & 0 & 0\end{array}\right)\;,m_{1}\approx m_{2}\quad,\begin{array}{c}
\Delta_{S}=m_{2}^{2}-m_{1}^{2}\\
\Delta_{A}=m_{1}^{2}\end{array}\,.\label{eq:DiagnonalIH}\end{equation}

With\begin{equation}
\begin{array}{c}
\Delta_{S}=(7.9\pm0.3)\times10^{-5}\,\mathrm{eV^{2}}\\
\Delta_{A}=(2.5\pm0.25)\times10^{-3}\,\mathrm{eV^{2}}\end{array}\,,\label{eq:ExpDeltas}\end{equation}
\begin{equation}
\begin{array}{c}
s_{12}^{2}\equiv\sin^{2}\theta_{12}=0.30\pm0.03\\
s_{23}^{2}\equiv\sin^{2}\theta_{23}=0.50\pm0.08\end{array}\,,\qquad s_{13}^{2}\equiv\sin^{2}\theta_{13}\leq0.02\,,\quad90\%\mathrm{CL\,.}\label{eq:ExpMixings}\end{equation}

Thus, apart from the poorly known $s_{13}$ mixing (we just know it
is small) and the phases, $\delta$ and $\phi$, the mass matrix can
be partially reconstructed in terms of the $2$ known mass differences
and the 2 known mixing angles for each of the two cases. In particular,
we can immediately extract the matrix element responsible for ($0\nu2\beta)$
decays: 

Normal hierarchy\begin{equation}
\langle m_{\nu}^{NH}\rangle_{ee}=(\mathcal{M}_{\nu}^{NH})_{ee}=\sqrt{\Delta_{S}}c_{13}^{2}s_{12}^{2}e^{i\phi}+\sqrt{\Delta_{A}}s_{13}^{2}e^{-i2\delta}\,.\label{eq:2betaNH}\end{equation}
In this case%
\footnote{Notice the dependence on the Dirac phase $\delta$. This is a consequence
of our convention for Majorana phases. One could redefine phases and
make this quantity independent on $\delta$, but this will not affect
predictions or constraints on observables. %
}, given the previous values, it is clear that $(\mathcal{M}_{\nu}^{NH})_{ee}\lesssim0.003\,\mathrm{eV}$
and therefore difficult to see in ($0\nu2\beta)$ decay experiments.

Inverted hierarchy\begin{equation}
\langle m_{\nu}^{IH}\rangle_{ee}=(\mathcal{M}_{\nu}^{IH})_{ee}=\sqrt{\Delta_{A}+\Delta_{S}}c_{13}^{2}s_{12}^{2}e^{i\phi}+\sqrt{\Delta_{A}}c_{13}^{2}c_{12}^{2}\,.\label{eq:2betaIH}\end{equation}
In this case, unless a cancellation occurs between the two terms for
$e^{i\phi}=-1$, $(\mathcal{M}_{\nu}^{IH})_{ee}$ is naturally of
order $0.05\,\mathrm{eV}$ and, therefore, observable in planned ($0\nu2\beta)$
decay experiments.

Equation~\eqref{eq:MnuYukawas} gives the mass matrix $\mathcal{M}_{\nu}$
in terms of the parameters of the model -- the Yukawa couplings, the
scalar masses and the trilinear coupling --, we can thus try to fix
some parameters by matching the $\mathcal{M}_{\nu}$, obtained from
the neutrino oscillation parameters, to the calculated one. Since
the mass matrix is symmetric, in principle this gives $6$ equations.
However, one of them is trivially satisfied because, by construction,
both matrices already satisfy $\det(\mathcal{M}_{\nu})=0$. To choose
the remaining $5$ equations we will use that the eigenvector corresponding
to the $0$ eigenvalue is very simple; as $\det f=0$, there is an
eigenvector $\mathbf{a}$ of $f$ with zero eigenvalue $f\boldsymbol{\cdot\mathbf{a}}=0$,
$\mathbf{a}=(f_{\mu\tau},-f_{e\tau},f_{e\mu})$. Obviously, $\mathbf{a}$
will also be an eigenvector of $\mathcal{M}_{\nu}$ with zero eigenvalue
when expressed in terms of masses and mixings and, therefore, ${UD}_{\nu}U^{T}\mathbf{a}=0$
or, since $U$ is unitary, $D_{\nu}U^{T}\mathbf{a}=0$. This gives
us three equations, one of which is satisfied trivially because one
of the diagonal values of $D_{\nu}$ is zero. The other two equations
will allow us to express the ratios of couplings $f_{ij}$ just in
terms of mixing angles and phases%
\footnote{Therefore, the decay branching ratios of the scalar $h$ to the different
leptons are fixed by the mixing angles. This can probably be exploited~\cite{AristizabalSierra:2006gb}
at the ILC . %
}. Thus, in the NH case we have $(D_{\nu}^{NH})_{11}=0$ and \begin{eqnarray}
\left(U^{T}\mathbf{a}\right)_{2}=0\;\Rightarrow\;\frac{f_{e\tau}}{f_{\mu\tau}} & = & \tan\theta_{12}\frac{\cos\theta_{23}}{\cos\theta_{13}}+\tan\theta_{13}\sin\theta_{23}e^{-i\delta}\:,\nonumber \\
\left(U^{T}\mathbf{a}\right)_{3}=0\;\Rightarrow\;\frac{f_{e\mu}}{f_{\mu\tau}} & = & \tan\theta_{12}\frac{\sin\theta_{23}}{\cos\theta_{13}}-\tan\theta_{13}\cos\theta_{23}e^{-i\delta}\:.\label{eq:efesangulosN}\end{eqnarray}
These equations immediately tell us that the standard PMNS convention
of phases is not compatible with all $f_{ab}$ being real. However,
we can take a phase convention in which $f_{\mu\tau}$ is real and
positive and leave $f_{e\tau}$ and $f_{e\mu}$ complex with phases
fixed by eq.~(\ref{eq:efesangulosN}).

With values like $s_{12}^{2}\sim0.3$, $s_{23}^{2}\sim0.5$ and $s_{13}^{2}<0.02$,
the first term on the right hand side of eqs.~(\ref{eq:efesangulosN})
dominates and we get $f_{e\tau}\simeq f_{\mu\tau}/2\simeq f_{e\mu}$.
With this relation we can go back to the low energy bounds in table
\ref{tab:universality} and table \ref{tab:meg} and find that the
strongest constraints on the $f_{ij}$ couplings come from $\mu\rightarrow e\gamma$
(which strongly bounds $|f_{e\tau}f_{\mu\tau}|$) and tell us that
$|f_{ei}|\lesssim0.05(m_{h}/\mathrm{TeV})$ and $|f_{\mu\tau}|\lesssim0.1(m_{h}/\mathrm{TeV})$.

The equations corresponding to the inverted hierarchy case, $(D_{\nu}^{IH})_{33}=0$,
are \begin{eqnarray}
\left(U^{T}\mathbf{a}\right)_{1}=0\;\Rightarrow\;\frac{f_{e\tau}}{f_{\mu\tau}} & = & -\frac{\sin\theta_{23}}{\tan\theta_{13}}e^{-i\delta}\:,\nonumber \\
\left(U^{T}\mathbf{a}\right)_{2}=0\;\Rightarrow\;\frac{f_{e\mu}}{f_{\mu\tau}} & = & \frac{\cos\theta_{23}}{\tan\theta_{13}}e^{-i\delta}\:.\label{eq:efesangulosI}\end{eqnarray}
 In this case, it is clear that $f_{e\tau}/f_{e\mu}=-\tan\theta_{23}\approx-1$
and $|f_{e\mu}|>5|f_{\mu\tau}$\textbar{}, $|f_{e\tau}|>5|f_{\mu\tau}|$.
Now we can use these relations in the low energy bounds in table \ref{tab:universality}
and table \ref{tab:meg} and find that the strongest constraints on
the $f_{ij}$ couplings come from lepton-hadron universality (see
table \ref{tab:universality}) , $5|f_{\mu\tau}|\lesssim|f_{ei}|\lesssim0.1(m_{h}/\mathrm{TeV})$.

We still have $3$ additional equations we can use to fix the parameters
of the model. In this case we have no good argument to choose them%
\footnote{Except that they cannot be in the same column or the same row of $\mathcal{M}_{\nu}$,
because in that case the equations are related by $\mathcal{M}_{\nu}\boldsymbol{\mathbf{a}}=0$.%
} and following \cite{Babu:2002uu} we will take the three elements
$m_{22}$, $m_{23}$ and $m_{33}$ in the equalities $m_{ij}\equiv\left(\mathcal{M}_{\nu}\right)_{ij}=\zeta f_{ia}m_{a}g_{ab}^{\ast}m_{b}f_{jb}$
with the $m_{ij}$ defined by eq. \eqref{eq:MnuUs} and $\zeta=\frac{\mu}{48\pi^{2}M^{2}}\tilde{I}$.
Thus, if  $\omega_{ab}=m_{a}g_{ab}^{*}m_{b}$ we have \begin{eqnarray}
m_{22} & = & \zeta(f_{\mu\tau}^{2}\omega_{\tau\tau}-2f_{e\mu}f_{\mu\tau}\omega_{e\tau}+f_{e\mu}^{2}\omega_{ee})\:,\nonumber \\
m_{23} & = & \zeta(-f_{\mu\tau}^{2}\omega_{\mu\tau}-f_{\mu\tau}f_{e\tau}\omega_{e\tau}+f_{\mu\tau}f_{e\mu}\omega_{e\mu}+f_{e\tau}f_{e\mu}\omega_{ee})\:,\nonumber \\
m_{33} & = & \zeta(f_{\mu\tau}^{2}\omega_{\mu\mu}+2f_{\mu\tau}f_{e\tau}\omega_{e\mu}+f_{e\tau}^{2}\omega_{ee})\:.\label{eq:mij}\end{eqnarray}
Because of the hierarchy among the charged lepton masses, it is natural
to assume that those $\omega_{ab}$ containing the electron mass,
$\omega_{ee},\,\omega_{e\mu},\,\omega_{e\tau}$, are much smaller
than $\omega_{\mu\mu},\,\omega_{\mu\tau},\,\omega_{\tau\tau}$, in
that case we can neglect them (we will check later the goodness of
this approximation within the numerical analysis), and we have \begin{equation}
m_{22}\simeq\zeta f_{\mu\tau}^{2}\omega_{\tau\tau}~,~m_{23}\simeq-\zeta f_{\mu\tau}^{2}\omega_{\mu\tau}~,~m_{33}\simeq\zeta f_{\mu\tau}^{2}\omega_{\mu\mu}~.\label{eq:mijApp}\end{equation}
 In the normal hierarchy case this gives ($s_{ij}\equiv\sin\theta_{ij}$,
$c_{ij}\equiv\cos\theta_{ij}$) \begin{eqnarray}
\zeta f_{\mu\tau}^{2}\omega_{\tau\tau} & \simeq & m_{3}c_{13}^{2}s_{23}^{2}+m_{2}e^{i\phi}(c_{12}c_{23}-e^{i\delta}s_{12}s_{13}s_{23})^{2}\:,\nonumber \\
\zeta f_{\mu\tau}^{2}\omega_{\mu\tau} & \simeq & -m_{3}c_{13}^{2}c_{23}s_{23}+m_{2}e^{i\phi}(c_{12}s_{23}+e^{i\delta}c_{23}s_{12}s_{13})(c_{12}c_{23}-e^{i\delta}s_{12}s_{13}s_{23})\:,\nonumber \\
\zeta f_{\mu\tau}^{2}\omega_{\mu\mu} & \simeq & m_{3}c_{13}^{2}c_{23}^{2}+m_{2}e^{i\phi}(c_{12}s_{23}+e^{i\delta}c_{23}s_{12}s_{13})^{2}.\label{eq:mijNangulos}\end{eqnarray}
With $m_{3}\simeq0.05$ eV and $m_{2}\simeq0.009$ eV, \begin{equation}
|\omega_{\tau\tau}|\simeq|\omega_{\mu\tau}|\simeq|\omega_{\mu\mu}|\simeq\frac{0.05\,\mathrm{eV}}{2\zeta|f_{\mu\tau}|^{2}}\,,\label{eq:omegasNH}\end{equation}
 setting a definite hierarchy among the $g_{ab}$ couplings: \begin{equation}
g_{\tau\tau}:g_{\mu\tau}:g_{\mu\mu}\sim m_{\mu}^{2}/m_{\tau}^{2}:m_{\mu}/m_{\tau}:1\:.\label{couplinghierarchy}\end{equation}

In the inverted hierarchy case, eqs. (\ref{eq:mij}) give \begin{eqnarray}
\zeta f_{\mu\tau}^{2}\omega_{\tau\tau} & \simeq & m_{1}(c_{23}s_{12}+e^{i\delta}c_{12}s_{13}s_{23})^{2}+m_{2}e^{i\phi}(c_{12}c_{23}-e^{i\delta}s_{12}s_{13}s_{23})^{2}\,,\nonumber \\
\zeta f_{\mu\tau}^{2}\omega_{\mu\tau} & \simeq & m_{1}(s_{12}s_{23}-e^{i\delta}c_{12}c_{23}s_{13})(c_{23}s_{12}+e^{i\delta}c_{12}s_{13}s_{23})\nonumber \\
 & + & m_{2}e^{i\phi}(c_{12}s_{23}+e^{i\delta}c_{23}s_{12}s_{13})(c_{12}c_{23}-e^{i\delta}s_{12}s_{13}s_{23})\,,\label{eq:mijIangulos}\\
\zeta f_{\mu\tau}^{2}\omega_{\mu\mu} & \simeq & m_{1}(s_{12}s_{23}-e^{i\delta}c_{12}c_{23}s_{13})^{2}+m_{2}e^{i\phi}(c_{12}s_{23}+e^{i\delta}c_{23}s_{12}s_{13})^{2}\,,\nonumber \end{eqnarray}
where $m_{1}\simeq m_{2}\simeq0.05$ eV, also yielding for $e^{i\phi}=1$
\begin{equation}
|\omega_{\tau\tau}|\simeq|\omega_{\mu\tau}|\simeq|\omega_{\mu\mu}|\simeq\frac{0.05\:\text{eV}}{2\zeta|f_{\mu\tau}|^{2}}\,,\label{eq:omegasIH}\end{equation}
and the hierarchy of couplings in eq.~(\ref{couplinghierarchy}).
However, in the IH case there is a strong cancellation for Majorana
phases close to $\pi$, and one can obtain a smaller value for $\omega_{\mu\mu}$,
thus we can only write\begin{equation}
|\omega_{\mu\mu}|>\frac{0.007\:\text{eV}}{2\zeta|f_{\mu\tau}|^{2}}\,.\end{equation}
In both cases one expects $g_{\mu\mu}$ to be the largest coupling
among the three considered. Of course, $g_{ee}$, $g_{e\mu}$ and
$g_{e\tau}$ can also be large and are only constrained by low energy
processes and perturbativity constraints. One should notice, however,
that in the inverted hierarchy case, the approximation made in going
from eqs.~(\ref{eq:mij}) to eqs.~(\ref{eq:mijApp}) may be a priori
less justifiable than in the normal hierarchy case when $\theta_{13}\to0$,
as the eigenvector corresponding to the zero eigenvalue, $(f_{\mu\tau},\,-f_{e\tau},\, f_{e\mu})$
is proportional to $(e^{i\delta}\tan\theta_{13},\,\sin\theta_{23},\,\cos\theta_{23})$,
i.e. $f_{\mu\tau}\propto\tan\theta_{13}$, and since the terms retained
in eq.~\eqref{eq:mij} are proportional to $f_{\mu\tau}$, it is
not obvious that the terms proportional to $\omega_{ei}$ can be neglected.

Assuming then that $|g_{\mu\tau}|\approx|g_{\mu\mu}|(m_{\mu}/m_{\tau})$
and $|g_{\tau\tau}|\approx|g_{\mu\mu}|(m_{\mu}/m_{\tau})^{2}$, we
can go back to tables \ref{tab:meee} and \ref{tab:meg} to find the
relevant constraints on the couplings. The best constraint comes from
$\tau^{-}\rightarrow\mu^{+}\mu^{-}\mu^{-}$, which tells us that $|g_{\mu\mu}|\lesssim0.4(m_{k}/\mathrm{TeV)}$,
$|g_{\mu\tau}|\lesssim0.024(m_{k}/\mathrm{TeV)}$, $|g_{\tau\tau}|\lesssim0.0015(m_{k}/\mathrm{TeV)}$. 

We can use all this information to set analytical bounds on the relevant
parameters of the model in the line discussed at the beginning of
section~\ref{secmas}.

\subsection{Analytical constraints\label{sub:Analytical-constraints}}

\subsubsection{NH case}

First, just from the neutrino mass formula, we have

\begin{equation}
|g_{\mu\mu}||f_{\mu\tau}|^{2}\geq10^{-3}\frac{\max(m_{k},m_{h})}{\tilde{I}\,\mathrm{TeV}}\frac{\max(m_{k},m_{h})}{\mathrm{\mu}}\,.\label{eq:lowerbound-gmm}\end{equation}
Now we can show that due to the logarithmic growth of $\tilde{I}$
for $m_{k}\gg m_{h}$ and the fact that $\tilde{I}\leq1$ for $m_{k}<m_{h}$, 

\begin{equation}
\frac{\max(m_{k},m_{h})}{\widetilde{I}m_{h}}\geq1\,.\end{equation}
Thus\begin{equation}
|g_{\mu\mu}||f_{\mu\tau}|^{2}\geq10^{-3}\frac{m_{h}}{\mathrm{TeV}}\frac{\max(m_{k},m_{h})}{\mathrm{\mu}}\,.\label{eq:lowerbound-gmm2}\end{equation}
Now we use the perturbativity bound on $\mu$, $\mu<\kappa\min(m_{h},m_{k})$\begin{equation}
\kappa|g_{\mu\mu}||f_{\mu\tau}|^{2}\geq10^{-3}\frac{m_{h}}{\mathrm{TeV}}\frac{\max(m_{k},m_{h})}{\min(m_{k},m_{h})}\,,\label{eq:master-bound}\end{equation}
which can be rewritten as (use that $m_{h}m_{k}=\max(m_{k},m_{h})\min(m_{k},m_{h})$)
\begin{equation}
\left(\frac{\max(m_{k},m_{h})}{\mathrm{TeV}}\right)^{2}\leq10^{3}\kappa|g_{\mu\mu}||f_{\mu\tau}|^{2}\frac{m_{k}}{\mathrm{TeV}}\mbox{}\end{equation}
We can use that $m_{k}\leq\max(m_{k},m_{h})$ and the perturbative
constraints $|g_{\mu\mu}|<\kappa$, $|f_{\mu\tau}|<\kappa$ to find
an upper limit on the masses of the charged scalars \begin{equation}
m_{h},\, m_{k}\leq\max(m_{k},m_{h})<10^{3}\kappa^{4}\,\mathrm{TeV}\,.\end{equation}
On the other hand, if we use $|g_{\mu\mu}|\lesssim0.4(m_{k}/\mathrm{TeV)}$,
coming from $\tau\rightarrow3\mu$ and $|f_{\mu\tau}|\lesssim0.1(m_{h}/\mathrm{TeV})$,
coming from $\mu\rightarrow e\gamma$ we immediately obtain a lower
bound on the masses of the scalars \begin{equation}
m_{k},m_{h}>\min(m_{k,}m_{h})>\frac{0.51}{\sqrt{\kappa}}\,\mathrm{TeV}\,.\end{equation}
If we only use the $\tau\rightarrow3\mu$ constraint in \eqref{eq:master-bound}
we find a bound on the $|f_{\mu\tau}|$ coupling\begin{equation}
\kappa|f_{\mu\tau}|^{2}\geq2.6\times10^{-3}\left(\frac{m_{h}}{\mathrm{\min}(m_{k},m_{h})}\right)^{2}\,,\label{eq:master-bound2}\end{equation}
and using that $m_{h}>\mathrm{\min}(m_{k},m_{h})$ we find an absolute
limit on the coupling\begin{equation}
|f_{\mu\tau}|>\frac{0.051}{\sqrt{\kappa}}\,.\end{equation}
Thus, using either the experimental bounds and/or the perturbativity
bounds, we can also set upper and lower limits on the different couplings
$|g_{\mu\mu}|$ , $\mu$ and the interesting observables $BR(\mu\rightarrow e\gamma)$
and $BR(\tau\rightarrow3\mu)$. As discussed in section~\ref{sec:LHC}
the LHC will be able to find the doubly charged scalar of the model
$k^{++}$ as long as it is lighter than about $1\,\mathrm{TeV}$,
thus it is interesting to know what are the constraints on the parameters
of the model if $m_{k}<1\,\mathrm{TeV}$. Following the procedure
described above one can also put strong limits on the parameters of
the model adding this additional constraint. We collect all the limits
we obtain in table~\ref{tab:Normal-Hierarchy-analytical}. It is
important to remark the assumptions we use to obtain these bounds:
we assume that because the small electron mass, as compared with the
tau lepton and muon masses, $\omega_{ie}\simeq0$. We also take central
values for the measured oscillation parameters. For other experimental
information, bounds on branching ratios of rare processes, we use
90\% CL limits. Finally the dependence on the perturbativity constraints
is encoded in the parameter $\kappa$ and explicitly displayed.

\begin{table}
\begin{tabular}{|c|c|}
\hline 
General case&
$m_{k}<\,1\,\mathrm{TeV}$\tabularnewline
\hline
\hline 
\parbox[c][0.07\textheight]{0.4\columnwidth}{%
\[
\frac{0.51}{\sqrt{\kappa}}\,\mathrm{TeV}\leq m_{h},\, m_{k}<10^{3}\kappa^{4}\,\mathrm{TeV}\]
}%
&
\parbox[c][0.07\textheight]{0.4\columnwidth}{%
\[
\frac{0.51}{\sqrt{\kappa}}\,\mathrm{TeV}\leq m_{h},m_{k}<24\kappa^{3/2}\,\mathrm{TeV}\]
}%
\tabularnewline
\hline 
\parbox[c][0.06\textheight]{0.4\columnwidth}{%
\[
\frac{0.1}{\kappa}\,\mathrm{TeV}\,<\mu<10^{3}\kappa^{5}\,\mathrm{TeV}\]
}%
&
\parbox[c][0.06\textheight]{0.4\columnwidth}{%
\[
0.26\,\mathrm{TeV}<\mu<\kappa\,\mathrm{TeV}\]
}%
\tabularnewline
\hline 
\parbox[c][0.07\textheight]{0.4\columnwidth}{%
\[
\frac{0.051}{\sqrt{\kappa}}<|f_{\mu\tau}|<\kappa\,,\;\frac{0.01}{\kappa^{2}}\leq|g_{\mu\mu}|\leq\kappa\,\]
}%
&
\parbox[c][0.07\textheight]{0.4\columnwidth}{%
\[
\frac{0.051}{\sqrt{\kappa}}<|f_{\mu\tau}|<\kappa\,,\;\frac{0.1}{\kappa}\leq|g_{\mu\mu}|\leq\kappa\,\]
}%
\tabularnewline
\hline 
\parbox[c][0.06\textheight]{0.4\columnwidth}{%
\[
BR(\mu\rightarrow e\gamma)\ge1.0\times10^{-19}/\kappa^{12}\]
}%
&
\parbox[c][0.06\textheight]{0.4\columnwidth}{%
\[
BR(\mu\rightarrow e\gamma)\ge8\times10^{-13}/\kappa^{2}\]
}%
\tabularnewline
\hline 
\parbox[c][0.06\textheight]{0.4\columnwidth}{%
\[
BR(\tau\rightarrow3\mu)\ge1.5\times10^{-18}/\kappa^{12}\]
}%
&
\parbox[c][0.06\textheight]{0.4\columnwidth}{%
\[
BR(\tau\rightarrow3\mu)\ge2\times10^{-10}/\kappa^{4}\]
}%
\tabularnewline
\hline
\end{tabular}

\caption{Normal Hierarchy analytical constraints: we assume $\omega_{ie}\simeq0$
and central values for measured oscillation parameters. For other
experimental information we use 90\% CL. The dependence on the perturbativity
constraints is encoded in the parameter $\kappa$ and explicitly displayed.\label{tab:Normal-Hierarchy-analytical} }
\end{table}

\subsubsection{IH case}

The same kind of bounds can be obtained for the IH case with a few
remarks. In the IH case $f_{\mu\tau}$ is not the largest coupling
among the $f's$, since $|f_{\mu\tau}|\approx\sqrt{2}s_{13}|f_{e\mu}$\textbar{}
with $s_{13}$ small. Thus perturbativity bounds should be applied
to $f_{e\mu}$. In addition the best experimental limit is also on
$f_{e\mu}$, $|f_{e\mu}|<0.1(m_{h}/\mathrm{TeV})$. Then, it is convenient
to write the main equations in terms of $f_{e\mu}$ instead of $f_{\mu\tau}$.
Finally in the IH hierarchy case there is the possibility of cancellations    
for $\phi=\pi$ which allow for a slightly smaller $\omega_{\mu\mu}$.
We have in this case \begin{equation}
s_{13}^{2}|g_{\mu\mu}||f_{e\mu}|^{2}\geq7.3\times10^{-5}\frac{m_{h}}{\mathrm{TeV}}\frac{\max(m_{k},m_{h})}{\mathrm{\mu}}\,.\label{eq:IH-lowerbound-gmm}\end{equation}
Then we can repeat essentially the same arguments used for the NH,
together with the upper limit on $s_{13}^{2}$, $s_{13}^{2}<0.02$,
to obtain lower and upper limits on the masses of the scalars, $m_{h},m_{k}$,
on the coupling $|f_{e\mu}|$, which is related to $|f_{e\tau}|$
and $|f_{\mu\tau}|$, on the coupling $|g_{\mu\mu}|$, related to
$|g_{\mu\tau}|$ and $|g_{\tau\tau}|$, and on the trilinear coupling
$\mu$. In addition, since in the IH case there is a strong dependence
on $s_{13}^{2}$ we can also set a lower bound on it. As in the NH
case we also give the corresponding limits one would find under the
assumption that the double charged scalar $k^{++}$ is found at the
LHC and, therefore, has a mass smaller than $1\,\mathrm{TeV}$. We
summarize all the limits in table~\ref{tab:Inverted-Hierarchy-analytical}.

\begin{table}
\begin{tabular}{|c|c|}
\hline 
General case&
$m_{k}<\,1\,\mathrm{TeV}$\tabularnewline
\hline
\hline 
\parbox[c][0.07\textheight]{0.45\columnwidth}{%
\[
\frac{0.95}{\sqrt{\kappa}}\,\mathrm{TeV}<m_{k},m_{h}<274\kappa^{4}\,\mathrm{TeV}\]
}%
&
\parbox[c][0.07\textheight]{0.45\columnwidth}{%
\[
\frac{0.95}{\sqrt{\kappa}}\,\mathrm{TeV}\leq m_{h},m_{k}<11\kappa^{3/2}\,\mathrm{TeV}\]
}%
\tabularnewline
\hline 
\parbox[c][0.07\textheight]{0.45\columnwidth}{%
\[
\frac{0.36}{\kappa}\,\mathrm{TeV}\,<\mu<274\kappa^{5}\,\mathrm{TeV}\]
}%
&
\parbox[c][0.07\textheight]{0.45\columnwidth}{%
\[
0.9\,\mathrm{TeV}<\mu<\kappa\,\mathrm{TeV}\]
}%
\tabularnewline
\hline 
\parbox[c][0.07\textheight]{0.45\columnwidth}{%
\[
\frac{0.095}{\sqrt{\kappa}}<|f_{e\mu}|<\kappa\,,\;\frac{0.036}{\kappa^{2}}\leq|g_{\mu\mu}|\leq\kappa\,\]
}%
&
\parbox[c][0.07\textheight]{0.45\columnwidth}{%
\[
\frac{0.1}{\sqrt{\kappa}}<|f_{e\mu}|<\kappa\,,\;\frac{0.36}{\kappa}\leq|g_{\mu\mu}|\leq\kappa\,\]
}%
\tabularnewline
\hline 
\parbox[c][0.06\textheight]{0.46\columnwidth}{%
\[
BR(\mu\rightarrow e\gamma)\ge2\times10^{-18}/\kappa^{12}\]
}%
&
\parbox[c][0.06\textheight]{0.45\columnwidth}{%
\[
BR(\mu\rightarrow e\gamma)\ge1\times10^{-12}/\kappa^{2}\]
}%
\tabularnewline
\hline 
\parbox[c][0.06\textheight]{0.45\columnwidth}{%
\[
BR(\tau\rightarrow3\mu)\ge2\times10^{-16}/\kappa^{12}\]
}%
&
\parbox[c][0.06\textheight]{0.45\columnwidth}{%
\[
BR(\tau\rightarrow3\mu)\ge3\times10^{-8}/\kappa^{4}\]
}%
\tabularnewline
\hline 
\parbox[c][0.06\textheight]{0.45\columnwidth}{%
\[
0.0007/\kappa^{3}<s_{13}^{2}<0.02\]
}%
&
\parbox[c][0.06\textheight]{0.45\columnwidth}{%
\[
0.018/\kappa<s_{13}^{2}<0.02\]
}%
\tabularnewline
\hline
\end{tabular}

\caption{Inverse Hierarchy analytical constraints.\label{tab:Inverted-Hierarchy-analytical} }
\end{table}

\subsection{Numerical analysis\label{sub:Numerical-analysis}}

The information we obtained above is very useful; however, to obtain
it we have made use of different approximations: 

\begin{description}
\item [{a)}] We assumed that the $\omega_{ee}$, $\omega_{e\mu}$, $\omega_{e\tau}$
can be neglected in front of the other couplings. This approximation
is reasonable because these $\omega$'s are proportional to the electron
mass, $\omega_{ei}=m_{e}g_{ei}m_{i}$, which is much smaller than
the other two lepton masses. However, it could happen that, for some
reason, the $g_{ei}$ couplings are much larger than the others. It
is therefore important to perform a complete analysis without this
assumption. 
\item [{b)}] We took central values for the measured oscillation parameters. 
\item [{c)}] In the analytical limits we only used data from neutrino oscillations
and bounds from $\tau\rightarrow3\mu$ and $\mu\rightarrow e\gamma$
(or lepton/hadron universality in the IH case) together with the perturbativity
constraints. As discussed in section~\ref{sub:Low-energy-constraints}
there are many more experimental constraints that can affect the results
and should be taken into account. 
\end{description}
It is clear that the only way to analyze the model without those approximations
is by means of an exhaustive numerical exploration of the parameter
space of the model. The basic tool to achieve this goal will be the
use of Monte Carlo (MC) techniques; however, because of the large
number of independent parameters and their diverse relevance, straightforward
application of MC techniques is not sufficiently efficient and thus
some additional considerations and refinements will be required.

The crudest MC exploration of the available parameter space would
involve random generation of complete sets of 17 independent basic
parameters, calculation of the corresponding predictions for the observables
and finally an acceptation/rejection process in terms of the agreement
between those predictions and the appropriate experimental constraints.
Beside the large number of parameters to be considered, the relations
among them previously discussed render such a crude approach almost
hopeless.

Realizing that not all observables play an equal role, that is, some
of them are much more informative or constraining than others, we
can go one further step in the use of simple MC techniques: instead
of the simplest MC outlined above, we can construct a MC process devised
to automatically produce mass matrices in agreement with neutrino
oscillation experiments.

Knowing the masses and mixing angles, if we were to reconstruct the
mass matrix $\mnu$ using experimental input, the only missing ingredients
would be the Dirac phase $\phd$, the Majorana phase $\phm$ (see
eqs.~(\ref{eq:U}--\ref{eq:DiagnonalIH})) and the poorly known mixing
$\theta_{13}$, for which we only know it is small and ignore its
exact value or even if it is zero. Equation~\eqref{eq:MnuYukawas}
gives the mass matrix $\mnu$ in terms of the new parameters -- the
Yukawa couplings, the scalar masses and the trilinear coupling --,
we can thus try to fix some parameters by matching the extracted $\mnu$
from oscillation data and the calculated one. This procedure achieves
two goals: it guarantees that neutrino oscillations are adequately
produced and it reduces the freedom in parameter space entering numerical
study by trading some of the couplings by measured neutrino oscillation
parameters. For each set $\{\theta_{12},\theta_{13},\theta_{23},\Delta_{A},\Delta_{S},\phd,\phm\}$
we thus obtain numerical values for the entries in $\mnu$. We will
then use eqs.~\eqref{eq:efesangulosN} or eqs.~\eqref{eq:efesangulosI}
to fix $\fem$ and $\fet$ in terms of $\fmt$ and the generated mixing
angles. Then come eqs.~(\ref{eq:mij}); these three complex relations
involve the six complex couplings $g_{ab}$, the trilinear coupling
and the scalar masses. Together with $m_{k}$, $m_{h}$ and $\mu$,
knowing three independent $g_{ab}$'s in eqs. (\ref{eq:mij}) will
be sufficient to fix the remaining ones; effectively this means that
we will generate $\gee$, $\gem$ and $\get$, and thus automatically
fix $\gmm$, $\gmt$ and $\gtt$. Notice that this phase convention
is compatible with the standard choice for the neutrino mass matrix,
eqs.~(\ref{eq:MnuUs}-\ref{eq:U}).\\
 To summarize, by generating five quantities -- $\Delta_{A}$, $\Delta_{S}$,
$\theta_{ij}$ -- according to experimental knowledge, two phases
-- $\phd$ and $\phm$ --, one real coupling $\fmt$, two masses --
$m_{h}$ and $m_{k}$ --, the trilinear coupling $\mu$ and three
complex $g_{ab}$, we are spanning the 12 moduli and 5 phases needed
to describe the model. That is, instead of the crude and utterly inefficient
Monte Carlo procedure in terms of $\{ m_{k},m_{h},\mu,f_{ab},g_{ab}\}$,
we can use $\{ s_{ij}^{2},\Delta_{A},\Delta_{S},\phd,\phm,\fmt,m_{h},m_{k},\mu,\gee,\gem\,\get\}$
to explore the whole parameter space and guarantee the agreement with
neutrino oscillations results prior to the use of the remaining experimental
constraints, which constitute the next step, as they are then applied
to accept/reject {}``candidate points''. Notice that we have not
specified the generation process of the different quantities involved:
some discussion will be addressed below, the details of the numerical
generation are summarized in table \ref{tab:values}.

\begin{table}[ht]
\begin{centering}\begin{tabular}{|c|c|c||c|c|c|}
\hline 
Parameter &
Value &
Shape &
Parameter &
Value &
Shape\tabularnewline
\hline 
$\Delta_{S}$ &
$(7.9\pm0.3)\times10^{-5}\,\mathrm{eV^{2}}$&
Flat &
$\Delta_{A}$ &
$(2.5\pm0.25)\times10^{-3}\,\mathrm{eV^{2}}$&
Flat\tabularnewline
\hline 
$\sin^{2}\theta_{12}$ &
$0.30\pm0.03$&
Flat &
$\sin^{2}\theta_{23}$ &
$0.50\pm0.08$&
Flat\tabularnewline
\hline 
$\sin^{2}\theta_{13}$ &
$[10^{-7};2\times10^{-2}]$&
Log flat &
&
&
\tabularnewline
\hline
\hline 
$\delta$ &
$[0;2\pi[$ &
Flat &
$\phi$ &
$[0;2\pi[$ &
Flat \tabularnewline
\hline 
$m_{h}$ &
$[10^{2};10^{9}]\,\mathrm{GeV}$&
Log Flat &
$m_{k}$ &
$[10^{2};10^{9}]\,\mathrm{GeV}$&
Log Flat \tabularnewline
\hline 
$f_{\mu\tau}$ &
$[10^{-7};\kappa]$&
Log flat &
$\mu$ &
$[1;10^{10}]\,\mathrm{GeV}$&
Log flat \tabularnewline
\hline 
$|g_{ee}|$ &
$[10^{-7};\kappa]$&
Log flat &
$\arg(g_{ee})$ &
$[0;2\pi[$&
Flat \tabularnewline
\hline 
$|g_{e\mu}|$ &
$[10^{-7};\kappa]$&
Log flat &
$\arg(g_{e\mu})$ &
$[0;2\pi[$&
Flat \tabularnewline
\hline 
$|g_{e\tau}|$ &
$[10^{-7};\kappa]$&
Log flat &
$\arg(g_{e\tau})$ &
$[0;2\pi[$&
Flat \tabularnewline
\hline
\end{tabular}\par\end{centering}

\caption{Numerical values.}

\label{tab:values} 
\end{table}

Despite being operative and useful, this refined MC procedure is not
the last word as one can do better. For this purpose we resort to
the use of Markov Chain driven Monte Carlo (MCMC) processes of the
Metropolis type.

We have discussed the benefits of a refined simple MC procedure with
respect to the crudest one: the next (and final) step to complete
the numerical toolkit we use is the rather straightforward conversion
of this refined MC into a Metropolis-like simulation which provides
the results to be discussed. This is largely beneficial as (1) the
efficiency of the MCMC process is sufficient to produce a reliable
and smooth output for the different subcases under study, (2) the
refined MC gives a helpful check of the consistency of the whole process.

Let us now discuss the remaining details concerning the simulations;
notice that, even if in the following we refer to the \emph{generation}
of parameters, which is appropriate for the MC process, the corresponding
feature when dealing with MCMC processes is not \emph{generation}
but in fact how they enter the stepwise acceptance function, however,
to avoid essentially duplicated discussions we will just mention what
concerns the plain MC case. The main idea that drives our election
of shapes and ranges of the different parameters is the need to perform
an adequate exploration of the available parameter space, in particular
one has to ensure that the regions which can yield interesting signals
like the production of scalars at the LHC or branching ratios of exotic
processes close to present experimental bounds are properly studied.
In particular: 

\begin{itemize}
\item Neutrino oscillations results, i.e. the squared masses differences
$\Delta_{A}$, $\Delta_{S}$ and the mixing parameters $\sin^{2}\theta_{12}$,
$\sin^{2}\theta_{23}$, are generated with flat distributions within
a $\pm1.64\sigma$ range around the quoted experimental value (this
corresponds to 90\% confidence level or probability region for a gaussian-distributed
uncertainty of the measurement). For $\sin^{2}\theta_{13}$, however,
we only have an upper bound: to span a reasonable range of values
it is generated through a logarithmically flat distribution from the
upper bound down to very small values, cut off at $10^{-7}$. 
\item The Dirac and Majorana phases, $\phd$ and $\phm$, are generated
according to flat distributions spanning the whole available range
$[0;2\pi[$. 
\item Concerning the independent Yukawa couplings $f_{ab}$ and $g_{ab}$,
moduli are generated through distributions logarithmically flat to
explore values that could potentially span several orders of magnitude.
The applied upper bounds correspond to the different naturalness/perturbative
cases under consideration. The arguments, as the Dirac and Majorana
phases, are generated through flat distributions over the complete
$[0;2\pi[$ range. 
\item The masses of the new scalar fields $m_{k}$ and $m_{h}$ are generated
with logarithmically flat distributions reaching up to $10^{5}$ TeV
and bounded below at $\sim100$ GeV to incorporate LEP-motivated constraints.
In any case, the precise upper bound is irrelevant as far as it is
beyond the analytic bounds presented in tables~\ref{tab:Normal-Hierarchy-analytical}
and \ref{tab:Inverted-Hierarchy-analytical}.
\item The trilinear coupling $\mu$ is also generated with a distribution
flat in its logarithm and limited by the perturbativity requirement. 
\item We apply the remaining experimental constraints presented in section~\ref{sub:Low-energy-constraints}
in a sharp (straightforward acceptance or rejection) way: the only
acceptable predictions are the ones within the quoted 90\% CL ranges/bounds.
\item The simulation described in the previous points allows a very wide
range of scalar masses. However, as discussed, the most interesting
case is when $m_{k}<1\,\mathrm{TeV}$ and, therefore, the $k^{++}$
can be discovered at the LHC. Thus, we have performed an independent
simulation requiring $m_{k}<1\,\mathrm{TeV}$.
\item All the simulations are done for both the NH and IH cases and for
two values of the perturbativity constraint $\kappa=1$ and $\kappa=5$.
\end{itemize}

The arbitrariness in the choice of priors and their impact in the
final results is always a concern in this type of analyses. Because of
this we have used several priors. In the case of the neutrino
oscillation parameters we have repeated the analysis fixing the
parameters at the central values, taking gaussian distributions around
central values and using the flat distributions we have finally
presented here. The differences are marginal and we chose to present
results for flat distributions because the results are slightly more
conservative. For other parameters we also tried plain flat priors,
but, specially for parameters that range in several orders of
magnitude, logarithmically flat distributions span more efficiently
the parameter space. We checked that the distributions obtained for
the observables considered, for which we found analytical lower and
upper limits, do not depend too much on the choice.

At this point the machinery used to perform the announced numerical
studies has been completely presented, however some comments on the
nature and interpretation of the output it produces are in order.

The model under study naturally {}``lives'' in a parametric space
of high dimensionality. The standard statistical arsenal offers two
different approaches to reduce this high dimensional information and
produce tolerably low dimensional -- usually one or two dimensional
-- output: the frequentist and the bayesian frameworks. Very schematically: 

\begin{itemize}
\item Frequentists assign confidence levels to the marginalized output through
the best fit achievable with the remaining parametric freedom. 
\item Bayesians assign probability densities to the marginalized output
through the integration over the remaining parametric freedom of the
likelihood (of data for the given parameters) times the prior distribution/weight
of parameters (this is just Bayes conditional probability inversion
formula at work). 
\end{itemize}
Beside the long standing quarrel existing among practitioners of one
or the other approach, both, with their information reduction schemes,
unavoidably present some drawbacks together with their statistical
merits. As we want sensitivity to the parameter space available for
the model to work, the procedure we have followed might look quite
bayesian. Being aware of the dependence on prior election and the
imprecise nature of the details behind many constraints%
\footnote{Usually the available information is just a 90\% CL range and little
additional knowledge on the distribution originating this range is
given. Moreover the perturbativity constraints, as clearly seen in
tables~\ref{tab:Normal-Hierarchy-analytical} and \ref{tab:Inverted-Hierarchy-analytical}
are determinant and, like all theoretical constraints, no obvious
confidence levels or statistical significance can be assigned to them. %
}, we do not intend at all to try and produce would-be highly orthodox
statistical results neither interpret them as if they were so, and
thus we have chosen the numerical details of the simulations -- that
is both experimental constraints and priors -- as stated above without
any further qualm.

\section{Results\label{sec:Results}}

In this section we collect the main results of the paper. 

First we would like to study the impact of the different assumptions
and experimental data in the analysis. 

To illustrate the impact of low energy constraints ($\mu\rightarrow e\gamma$,
$\tau\rightarrow3\mu$, ...) we perform an independent simulation
with only neutrino data and another simulation including all constraints
(in the case of IH and $\kappa=5$) and represent the resulting distribution%
\footnote{In the following we obtain the distributions as five million point
samples from a MCMC exploration of the parameter space as described
in section~\ref{sub:Numerical-analysis}.%
} of $m_{k}$ for the two simulations. We represent with a dashed line
the results of the simulation only with neutrino data and with a solid
line the simulation with all present experiments included. 
\vspace*{4mm}
\begin{figure}[ht]
\noindent \begin{centering}\includegraphics[width=0.55\columnwidth]{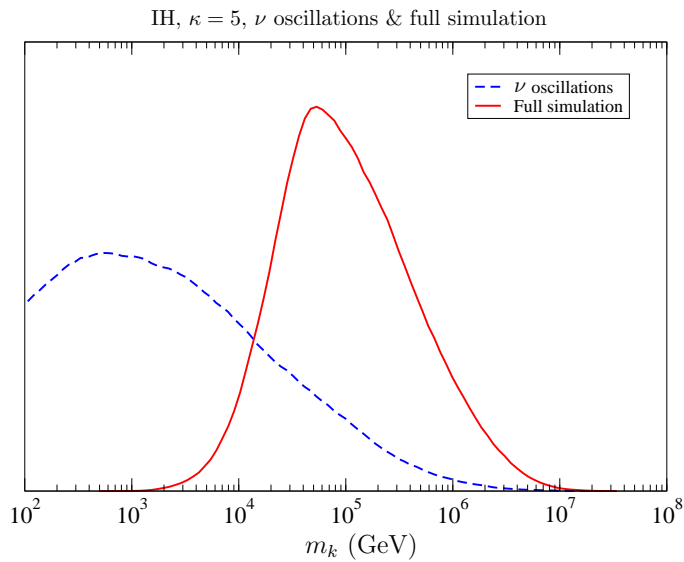}\par\end{centering}
\caption{\noindent Impact of low energy constraints ($\mu\rightarrow e\gamma,\tau\rightarrow3\mu$,
...): $m_{k}$ distribution when only neutrino data is included (dashed)
as compared with the case in which all experiments are included (solid).
Displayed data correspond to the IH case and $\kappa=5$.\label{fig:impact-low}}
\end{figure}

It is clear from the figure that only neutrino data allow (even prefer)
relatively low masses of the order of $1\,\mathrm{TeV}$ or below.
However, when low energy experimental data is included the lower limit
on the $m_{k}$ is pushed to larger values. We have to remark that
the shape of the curves basically reflects the volume of the parameter
space (from the other parameters) and that the tails can be rather
long, thus regions below $m_{k}<1\,\mathrm{TeV}$, as we will see
by performing simulations with $m_{k}<1\,\mathrm{TeV}$, are not completely
forbidden.

\begin{figure}[ht]
\noindent \begin{centering}\includegraphics[clip,width=0.55\columnwidth]{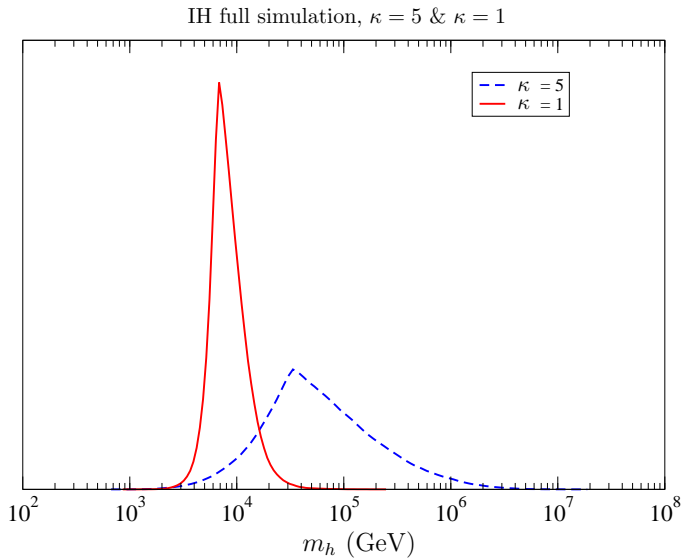}\par\end{centering}

\caption{\noindent Impact of the perturbativity constraints: $m_{h}$ distribution
for $\kappa=1$ (dashed) and $\kappa=5$ (solid). IH case with all
experimental data included.\label{fig:impact-kappa}}
\end{figure}

Similarly, to illustrate the impact of the perturbativity constraint,
we represent in fig.~\ref{fig:impact-kappa} the $m_{h}$ distribution 
from a simulation with $\kappa=1$
and another one with $\kappa=5$ (both in IH case with all experimental
information included).

We confirm with this figure the scaling of the bounds with the perturbativity
assumptions, encoded in the parameter $\kappa$, obtained analytically.
Thus, for smaller values of $\kappa$ the allowed range of $m_{h}$
is much smaller. For $\kappa=1$, the preferred region of $m_{h}$
is in the range $\sim1-100$~TeV (although with long tails) while
for $\kappa=5$ this range is enlarged to $1-10000$~TeV. Although
it cannot be appreciated in this figure the lower bound on the mass
is also sensitive to $\kappa$ as shown analytically (see table~\ref{tab:Inverted-Hierarchy-analytical}).

\begin{figure}[ht]
\begin{centering}\includegraphics[clip,width=0.55\columnwidth]{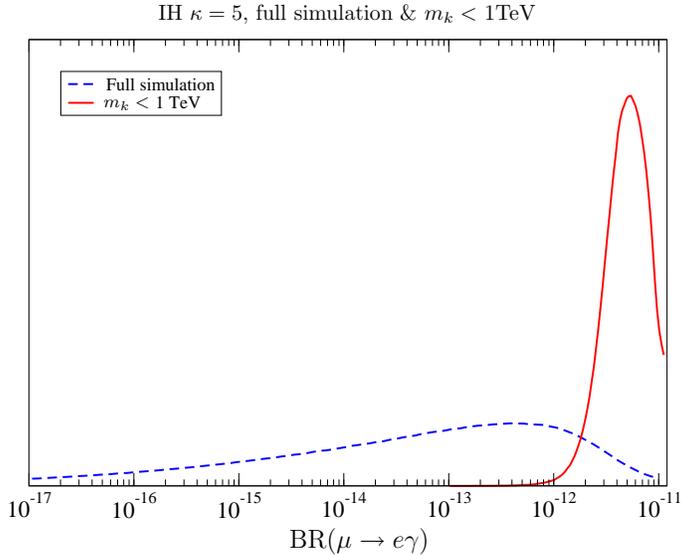}\par\end{centering}

\caption{\noindent Impact of the discovery of the $k^{++}$ at the LHC ($m_{k}<1\,\mathrm{TeV}$):
$BR(\mu\rightarrow e\gamma)$ distribution for the general case, IH
and $\kappa=5$ with all present experimental results, (dashed) and
requiring in addition that $k^{++}$ has been seen at the LHC ($m_{k}<1\,\mathrm{TeV}$)
$\kappa=5$ (solid). \label{fig:impact-lhc}}
\end{figure}

From figure~\ref{fig:impact-low} it is clear that all present data
allow a wide range of $k^{++}$ masses however, the $k^{++}$, as
discussed in \ref{sec:LHC}, can only be discovered at the LHC if
$m_{k}<1\,\mathrm{TeV}$. Thus, this is the really interesting region
of parameters to be studied. To study this region we perform an independent
simulation implementing (we present results for the IH case and $\kappa=5$)
all present constraints but assuming, in addition, that the $k^{++}$
has been discovered at the LHC and therefore has a mass $m_{k}<1\,\mathrm{TeV}$.
In fig.~\ref{fig:impact-lhc} we present the $BR(\mu\rightarrow e\gamma)$
distribution in the two cases, general case and $m_{k}<1\,\mathrm{TeV}$.
We see the dramatic impact in $BR(\mu\rightarrow e\gamma)$ of the
discovery of the $k^{++}$ at the LHC. While present data allow branching
ratios in the range $10^{-22}-10^{-11}$, if the $k^{++}$ is discovered
at the LHC then $BR(\mu\rightarrow e\gamma)\sim10^{-13}-10^{-11}$
and, therefore, will be probed at the MEG experiment.

\begin{figure}[ht]
\begin{centering}\includegraphics[clip,width=0.55\columnwidth]{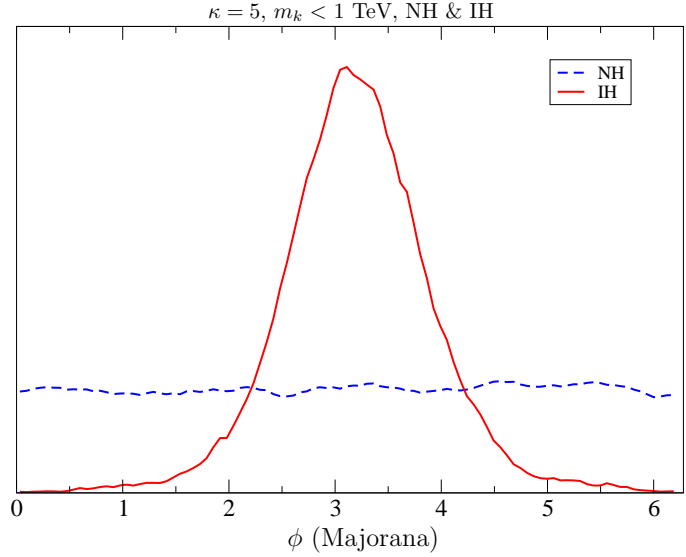} \par\end{centering}

\caption{\noindent Differences between the NH and IH cases: distribution of
the Majorana phase $\phi$: dashed in the NH case and solid in the
IH case. All present experimental data included and $\kappa=5$.\label{fig:impact-nhvsihphi}}
\end{figure}

Until now we have presented results only for the IH case. In general,
as also seen in our approximate analytical results, we expect roughly
similar results in the NH and the IH case, except for a few parameters
and/or observables. In particular we mentioned that in the IH case
there is a preference for the Majorana phase around $\phi=\pi$ because
in that case there is a cancellation in the neutrino mass formulas.
This is confirmed by the numerical calculation: in fig.~\ref{fig:impact-nhvsihphi}
we represent the distribution of $\phi$ for both, the NH case (dashed
line) and the IH case (solid line). The data is taken from a simulation
with $\kappa=5$, including all present experimental constraints and
requiring that $m_{k}<1$~TeV. The distribution in the NH case is
practically flat, while in the IH case it is highly peaked at $\phi=\pi$. 

\begin{figure}[ht]
\begin{centering}\includegraphics[clip,width=0.55\columnwidth]{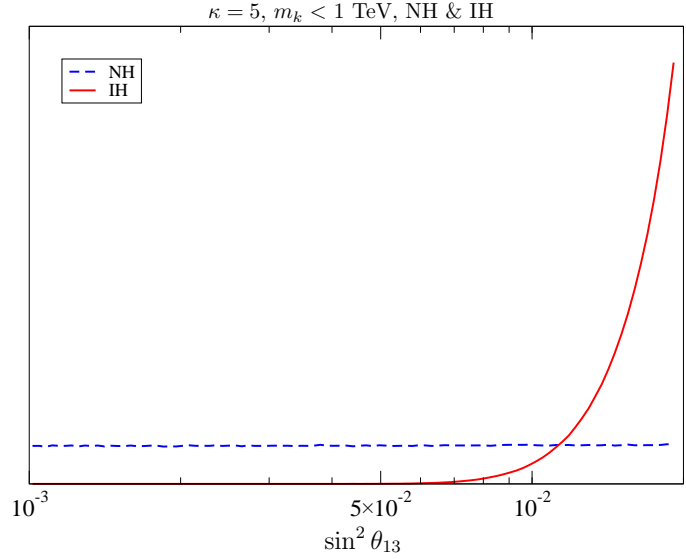}\par\end{centering}

\caption{\noindent Differences between the NH and IH cases: distribution of
$\sin^{2}\theta_{13}$ as in fig.~\ref{fig:impact-nhvsihphi}.\label{fig:impact-nhvsihs2th13}}
\end{figure}

We also expect large differences in the NH and IH cases for the parameter
$\sin^{2}\theta_{13}$. In fig.~ \ref{fig:impact-nhvsihs2th13} we
represent the $\sin^{2}\theta_{13}$ distribution for the two cases,
NH and IH ($\kappa=5$, full data and $m_{k}<1$~TeV). While it is
constant in the NH case, in the IH case it is highly peaked at the
maximum values allowed by present data and, in fact, there is an absolute
lower bound on it, $\sin^{2}\theta_{13}\gtrsim2\times10^{-3}$, which
is not so far from the present upper limit, $\sin^{2}\theta_{13}<2\times10^{-2}$.

\begin{figure}[ht]
\begin{centering}\includegraphics[clip,width=0.55\columnwidth]{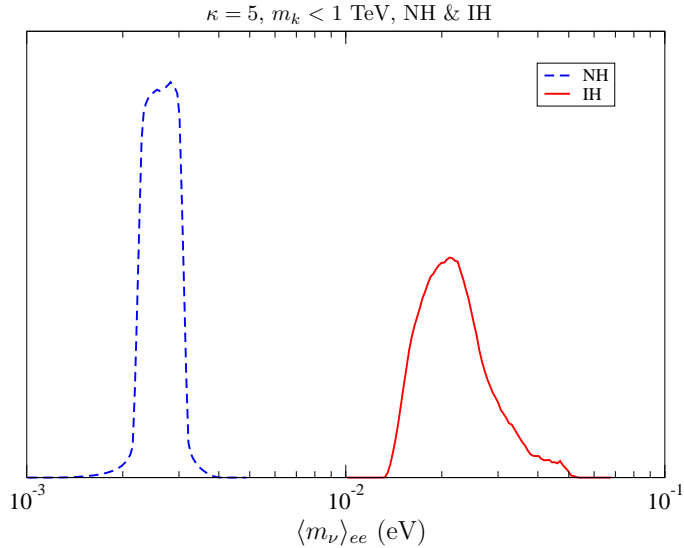}\par\end{centering}

\caption{\noindent Predictions for $\langle m_{\nu}\rangle_{ee}$ in the NH
(dashed) and the IH (solid) cases. All present experimental data included
and $\kappa=5$.\label{fig:impact-mnuee}}
\end{figure}

Finally, to illustrate another interesting difference between the
two cases, NH and IH, we have represented in fig.~\ref{fig:impact-mnuee}
the distribution of $\langle m_{\nu}\rangle_{ee}$, the relevant matrix
element in the neutrinoless double beta decay experiments. As before
we assume $\kappa=5$ and $m_{k}<1\,$TeV, however, this assumption
has little influence on the result since, as shown in eqs.~(\ref{eq:2betaNH}-\ref{eq:2betaIH}),
$\langle m_{\nu}\rangle_{ee}$ is a function of only the neutrino
masses and the mixing angles. Thus, the shape of the curves and their
position is just a consequence of the fact that the model predicts
a massless neutrino. In any case, from the figure it is clear that
the model predicts $\langle m_{\nu}\rangle_{ee}\sim0.001-0.005$~eV
in the NH and $\langle m_{\nu}\rangle_{ee}\sim0.01-0.06$~eV in the
IH case.

Now, for the most interesting observables, those which give some interesting
constraints or good perspectives in future tests, we present two-dimensional
contour plots of the corresponding distributions. The density of points has 
been calculated using 50 bins in a logarithmic scale for each axis. Then, 10 
contour lines equally spaced, ranging from the maximum density to 
$1/1000$ of it, have been represented. Thus, the last contour region,
painted with a lighter color, represents the region with a small density of 
points but which still contains some points.  
For each pair of observables we present two plots. On the left we
present the distribution when all present experimental constraints
are imposed. The values in the interesting region for the LHC ($m_{k}<1\,\mathrm{TeV}$)
are very low but not zero. Thus, to better study this region we present
in a second plot (right) the results of a simulation imposing the
additional constraint $m_{k}<1\,\mathrm{TeV}$. All results are given
for the most conservative perturbativity assumption ($\kappa=5$).
Scaling for more restrictive assumptions can be inferred from tables~\ref{tab:Normal-Hierarchy-analytical}
and \ref{tab:Inverted-Hierarchy-analytical}. We discuss the relevant
plots for both the NH and IH cases.

\subsection{Normal Hierarchy}

Here we consider correlations among observables in the normal hierarchy
case.

\begin{figure}[ht]
\noindent \begin{centering}\includegraphics[width=0.49\columnwidth]{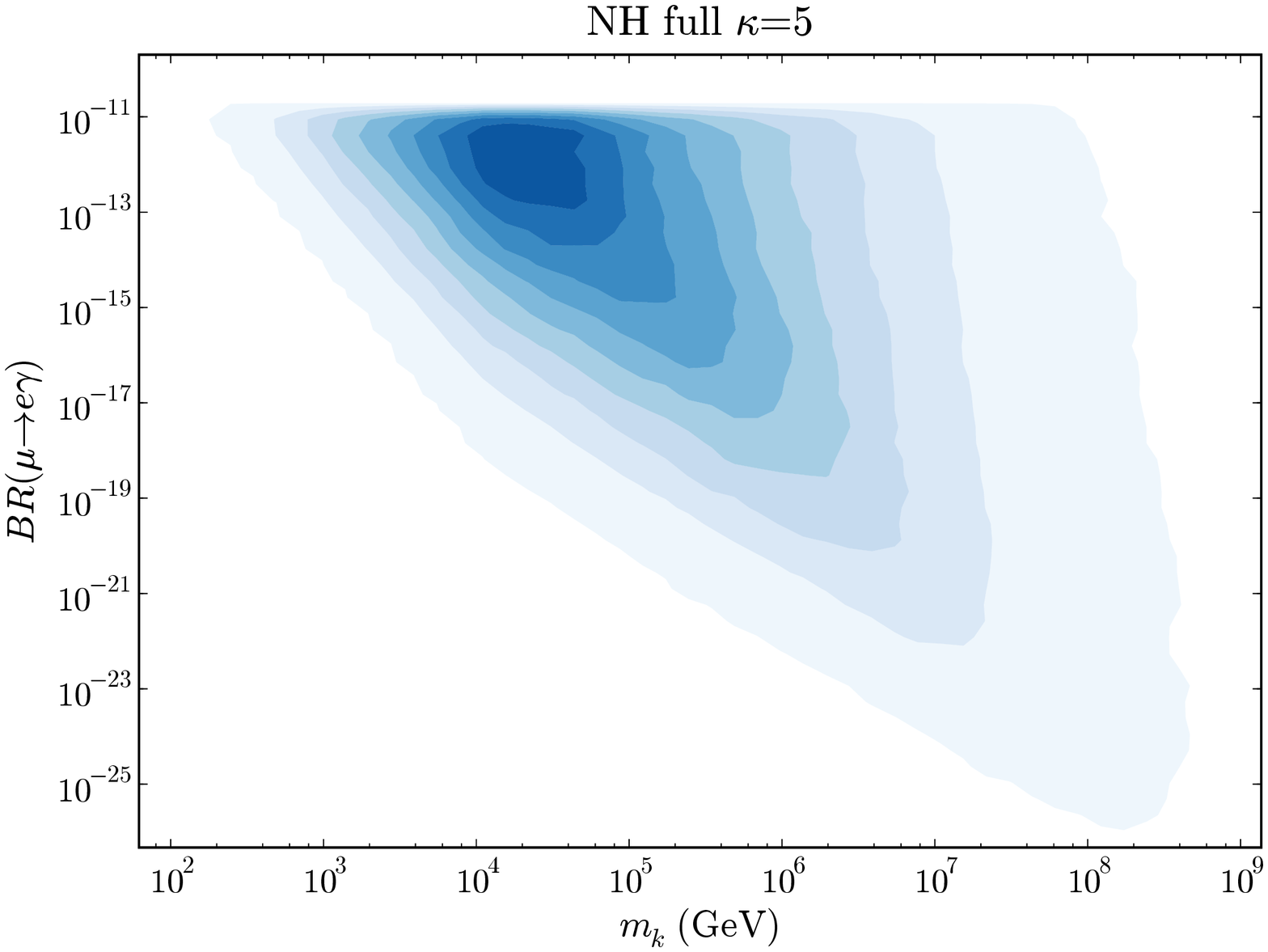}\includegraphics[width=0.49\columnwidth]{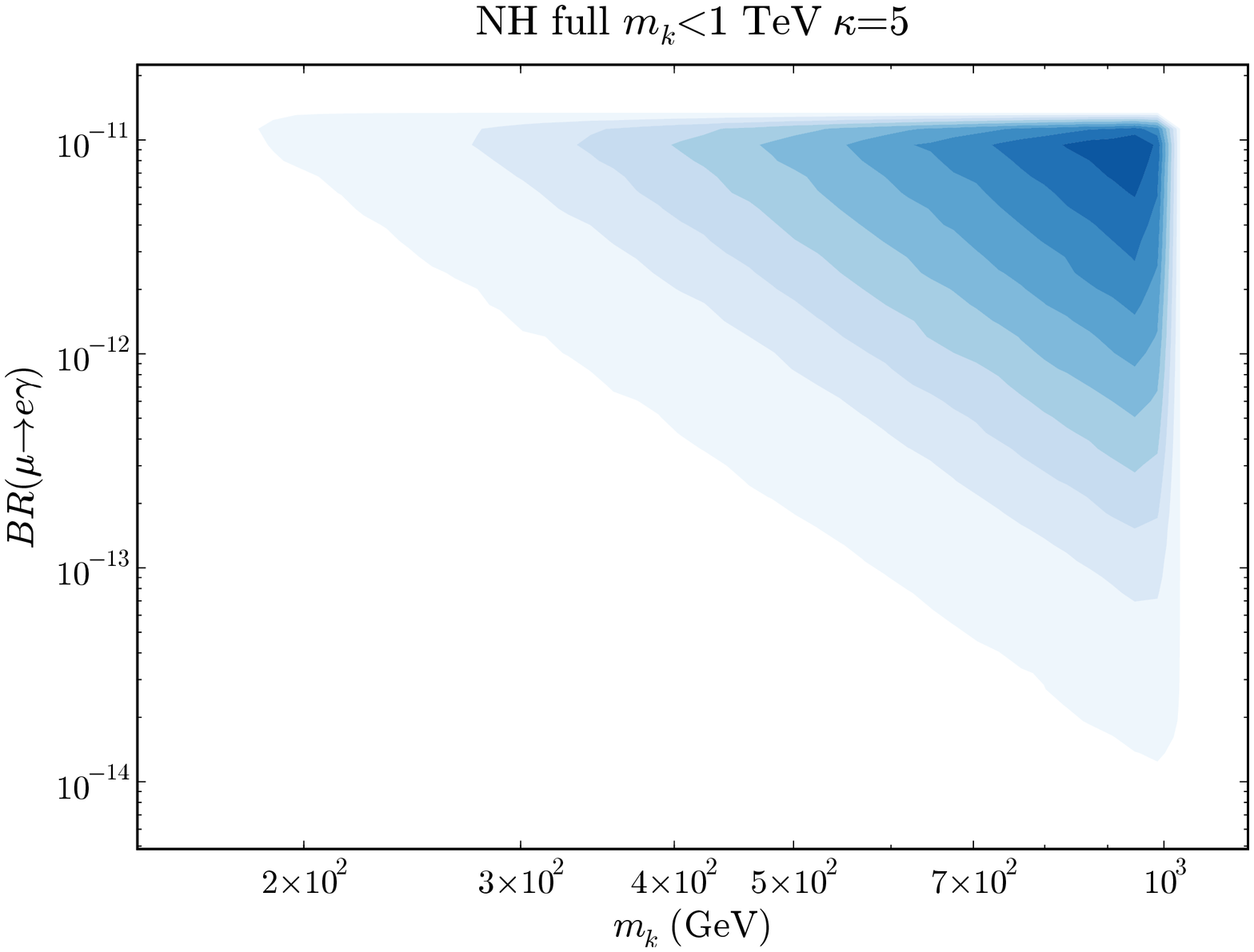}\par\end{centering}

\caption{\noindent NH: $BR(\mu\rightarrow e\gamma)$ vs $m_{k}$; (left) general
case (right) assuming that the $k^{++}$ is seen at the LHC ($m_{k}<1\,\mathrm{TeV}$).\label{fig:NH:m2eg-mk}}
\end{figure}

The most interesting observables of the model are $m_{k}$, because,
if small enough, it will be accessible at the LHC, and the $BR(\mu\rightarrow e\gamma)$,
which will be probed at a precision of the order of $10^{-13}$ in
the MEG experiment. Thus in fig.~\ref{fig:NH:m2eg-mk} we display
the joint $BR(\mu\rightarrow e\gamma)$-$m_{k}$ distribution. We
observe a clear correlation between these two observables. Present
data (left) seems to prefer $m_{k}\sim10\,\mathrm{TeV}$ and $BR(\mu\rightarrow e\gamma)$
above $10^{-13}$, but a large region of values is not excluded $m_{k}\sim10^{2}-10^{8}\,\mathrm{GeV}$
and $BR(\mu\rightarrow e\gamma)\sim10^{-25}-10^{-11}$, however if
the doubly charged scalar, $k^{++}$, is discovered at the LHC ($m_{k}<1\,\mathrm{TeV}$)
the situation changes dramatically and the simulation shows that the
preferred values are in the upper range%
\footnote{For specific numbers we take 3 contours in the plots.%
} $m_{k}\gtrsim 600\,\mathrm{GeV}$ and $BR(\mu\rightarrow e\gamma)\gtrsim10^{-12}$
and that $k^{++}$ masses below $200\,\mathrm{GeV}$ and $BR(\mu\rightarrow e\gamma)$
below $10^{-14}$ are very difficult to obtain in the model.

\begin{figure}[ht]
\noindent \begin{centering}\includegraphics[width=0.49\columnwidth]{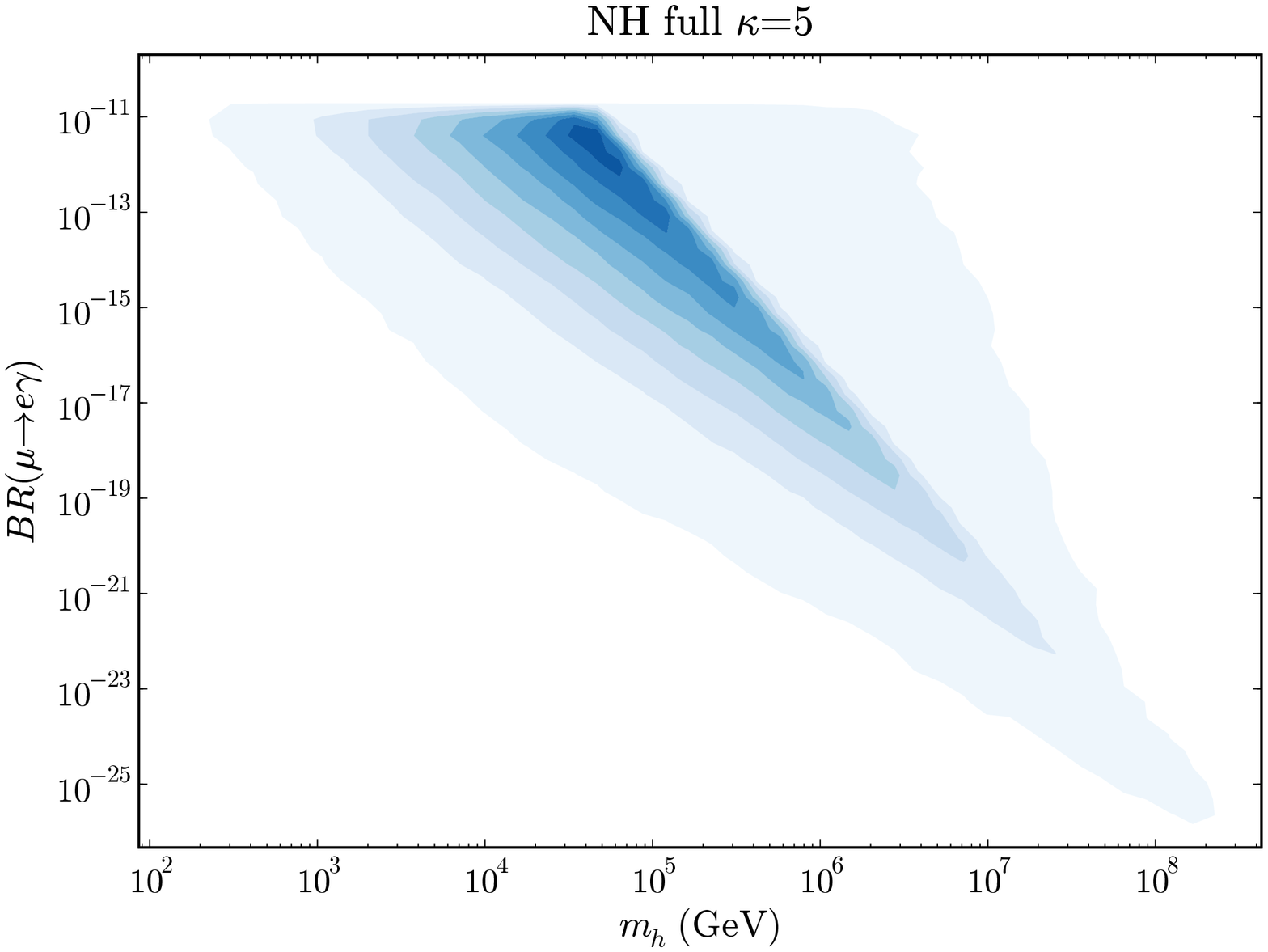}\includegraphics[width=0.49\columnwidth]{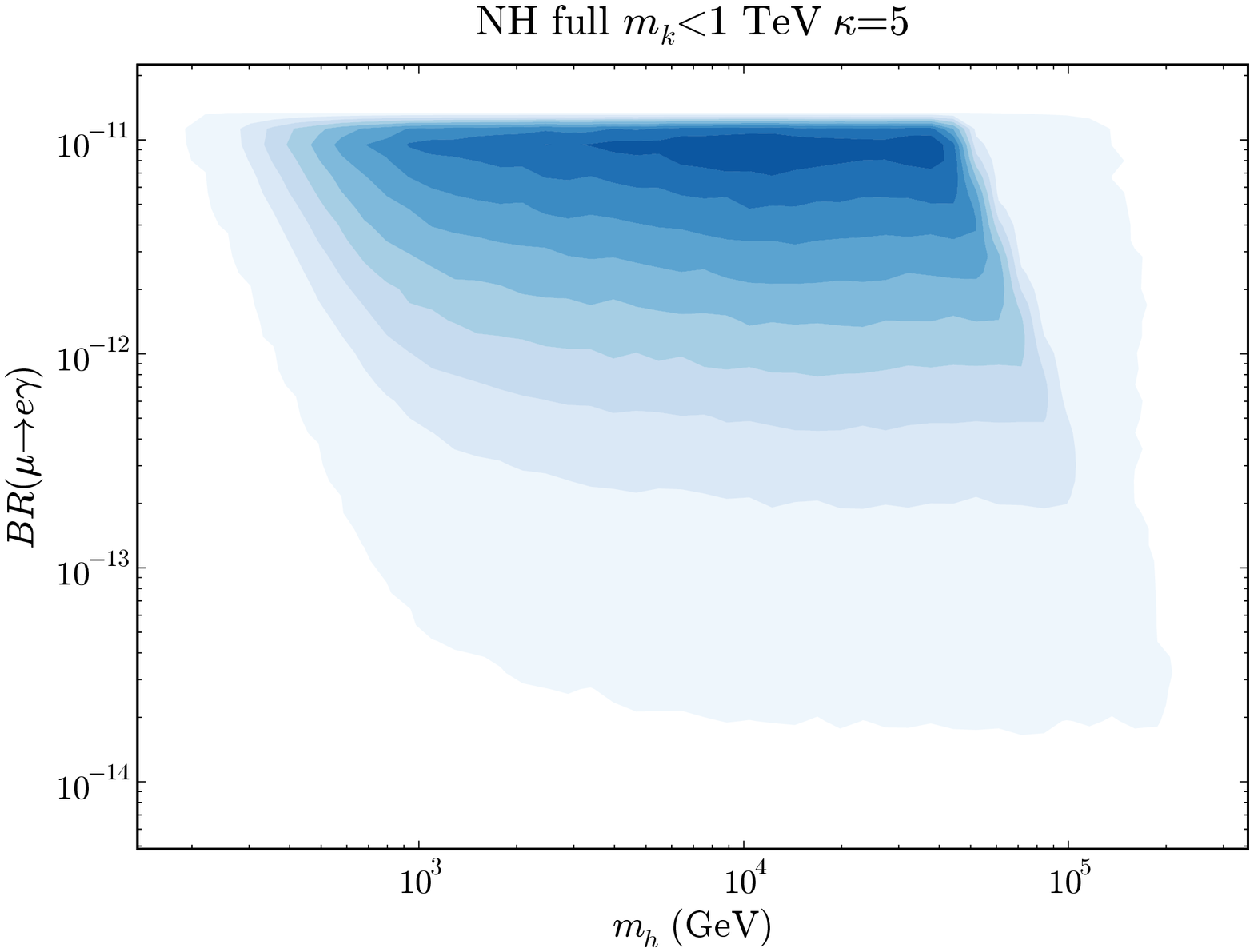}\par\end{centering}

\caption{\noindent NH: $BR(\mu\rightarrow e\gamma)$ vs $m_{h}$.\label{fig:NH:m2eg-mh}}
\end{figure}

\noindent Since $BR(\mu\rightarrow e\gamma)$ depends more explicitly
on $m_{h}$ than on $m_{k}$ it is interesting to study the correlation
between $BR(\mu\rightarrow e\gamma)$ and $m_{h}$. In fig.~\ref{fig:NH:m2eg-mh}
we depict the allowed region in the plane $BR(\mu\rightarrow e\gamma)$--$m_{h}$,
on the left for the general case and on the right for the case $m_{k}<1\,\mathrm{TeV}$.
We see a strong correlation specially in the general case. From the
figure on the left we see that in the general case $m_{h}$ can be
in a very wide range of values $m_{h}\sim10^{2}-10^{8}\,\mathrm{GeV}$
but the preferred values are $m_{h}\sim40\,\mathrm{TeV}$. On the
other hand, if $m_{k}<1\,\mathrm{TeV}$ the allowed range of $m_{h}$
is much smaller, $m_{h}\sim10^{2}-10^{5}\,\mathrm{GeV}$, and is shifted
to the lower edge. It still allows a large range of masses not accessible
at the LHC. 

\begin{figure}[ht]
\noindent \begin{centering}\includegraphics[width=0.49\columnwidth]{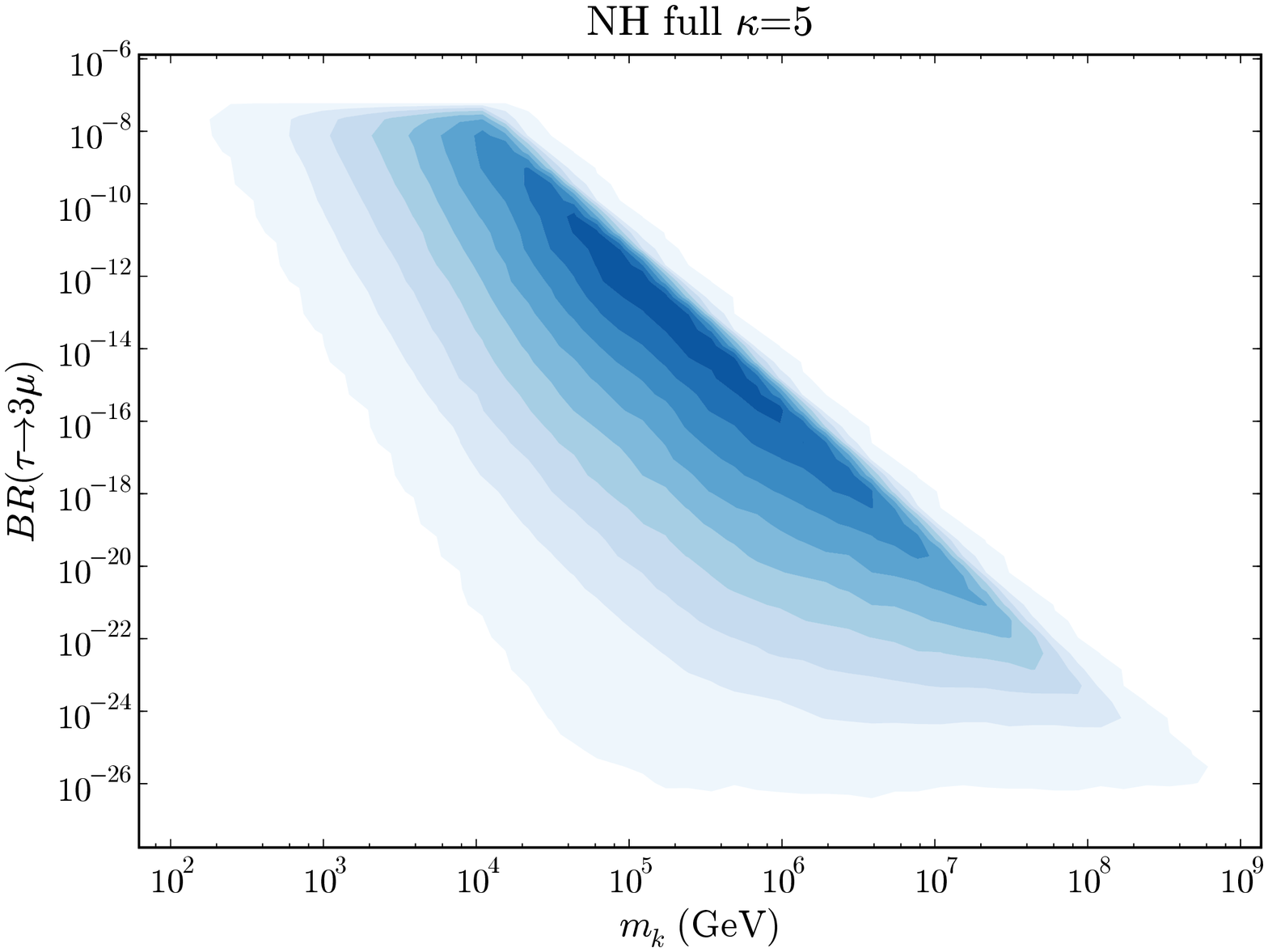}\includegraphics[width=0.49\columnwidth]{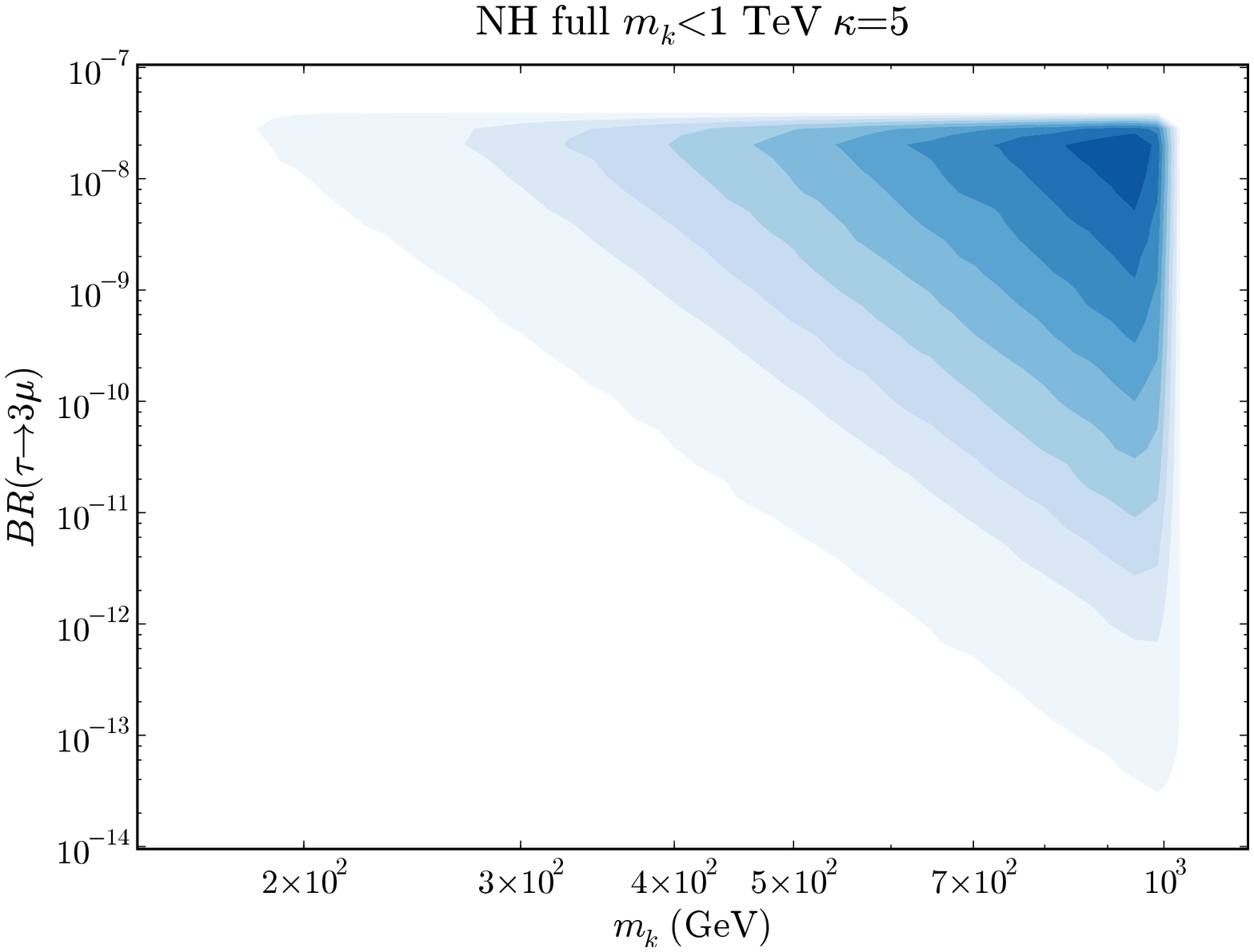}\par\end{centering}

\caption{\noindent NH: $BR(\tau\rightarrow3\mu)$ vs $m_{k}$.\label{fig:NH:t2mmm-mk}}
\end{figure}

\noindent $\tau\rightarrow3\mu$ is mediated by $k^{++}$ exchange
and governed by couplings which are largely fixed by neutrino mass
data. Thus in fig.~\ref{fig:NH:t2mmm-mk} we present results for
the allowed values in the plane $BR(\tau\rightarrow3\mu)$--$m_{k}$.
As expected there is a strong correlation in the two cases considered.
In the general case the preferred values of $BR(\tau\rightarrow3\mu)$
are in the $10^{-13}$ range, although values as small as $10^{-25}$
are allowed. In the $m_{k}<1\,\mathrm{TeV}$ case the preferred values
are $BR(\tau\rightarrow3\mu)\gtrsim10^{-9}$ which is not so far from
present limits, $BR(\tau\rightarrow3\mu)<3.2\times10^{-8}$ , but
values like $10^{-13}$ are not completely excluded. 

\subsection{Inverted Hierarchy}

In this section we consider correlations among observables in the
inverted hierarchy case.

\begin{figure}[ht]
\noindent \begin{centering}\includegraphics[width=0.49\columnwidth]{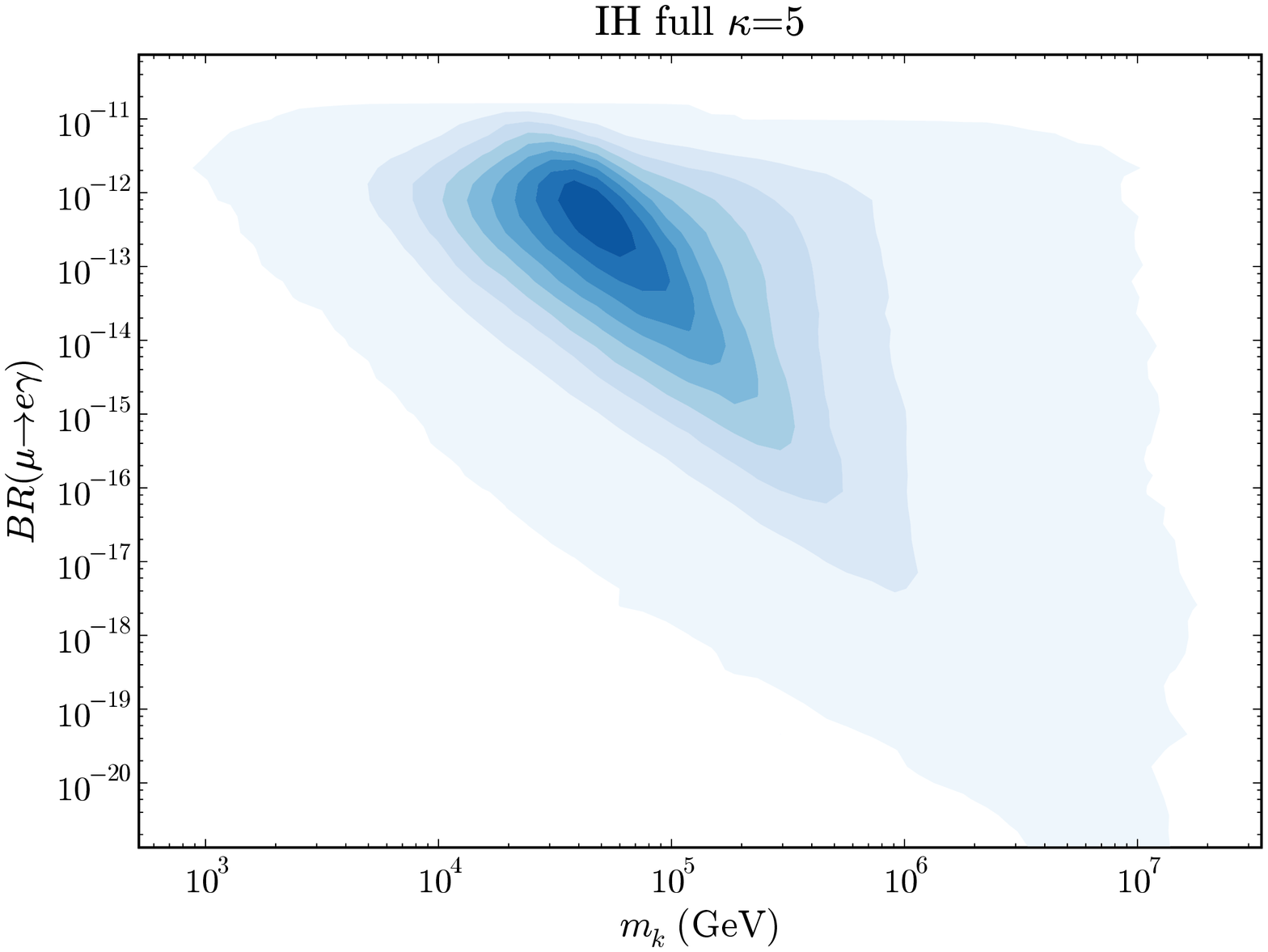}\includegraphics[width=0.49\columnwidth]{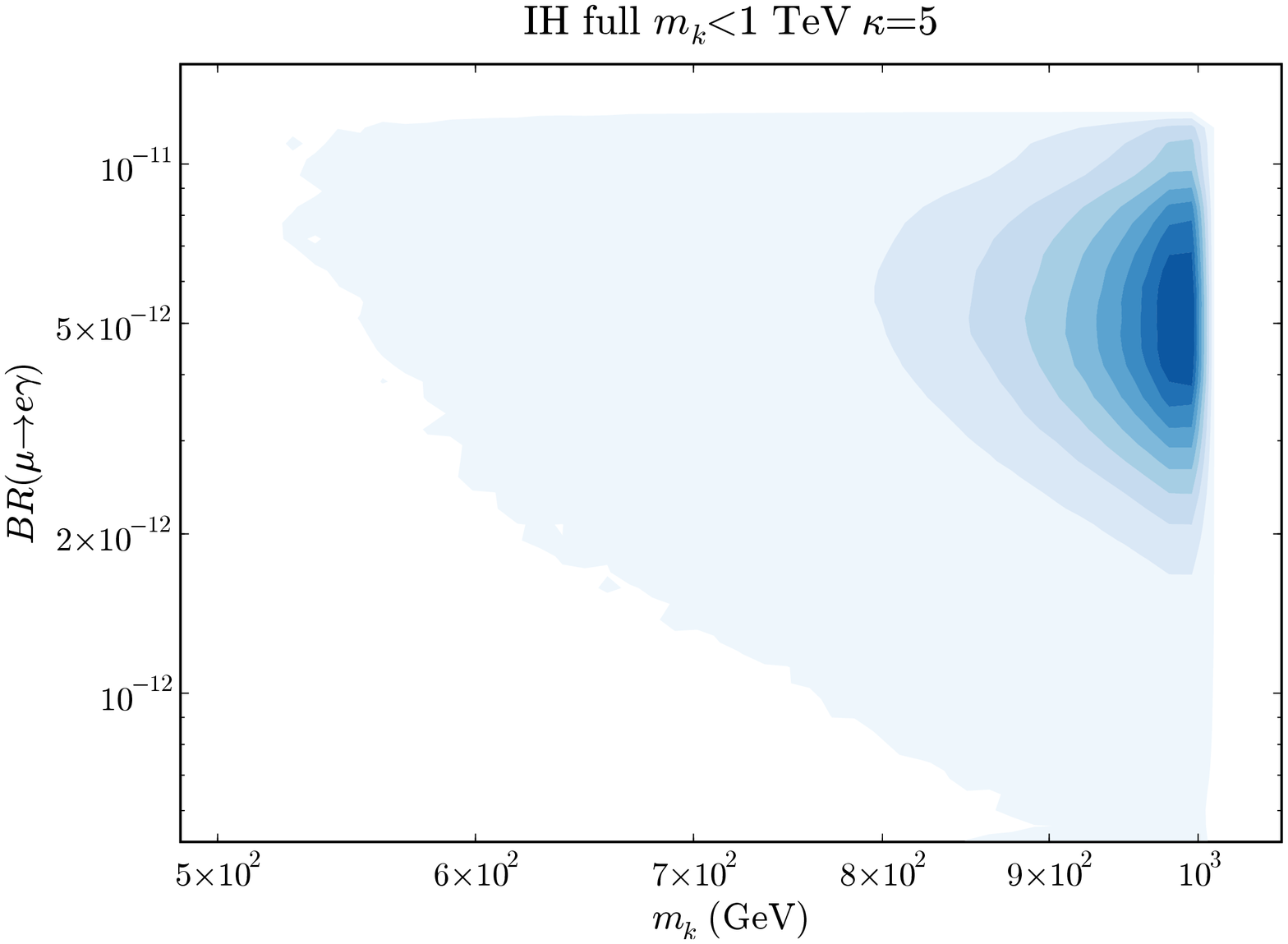}\par\end{centering}

\caption{\noindent IH: $BR(\mu\rightarrow e\gamma)$ vs $m_{k}$.\label{fig:IH:m2eg-mk}}
\end{figure}

\noindent In fig.~\ref{fig:IH:m2eg-mk} we represent $BR(\mu\rightarrow e\gamma)$
vs $m_{k}$. On the left for the general case and on the right with
the additional assumption that the $k^{++}$ has been seen at the
LHC ($m_{k}<1\,\mathrm{TeV}$). The plots are similar to the plots
obtained in the NH case, although slightly more restrictive. The allowed
region in the general case is $m_{k}\sim10^{2}-10^{7}\,\mathrm{GeV}$
and $BR(\mu\rightarrow e\gamma)\sim10^{-22}-10^{-11}$ and higher
density values occur for $m_{k}\sim50\,\mathrm{TeV}$ and $BR(\mu\rightarrow e\gamma)\sim10^{-12}$.
However, if nature chooses a $k^{++}$ light enough to be produced
at the LHC the model is much more constrained: it predicts that $m_{k}$
is relatively large (masses below $400\,\mathrm{GeV}$ are only marginally
allowed and the preferred masses are above $800\,\mathrm{GeV}$).
In addition $BR(\mu\rightarrow e\gamma)>10^{-13}$ and the preferred
range is above $2\times10^{-12}$.

\begin{figure}[ht]
\noindent \begin{centering}\includegraphics[width=0.49\columnwidth]{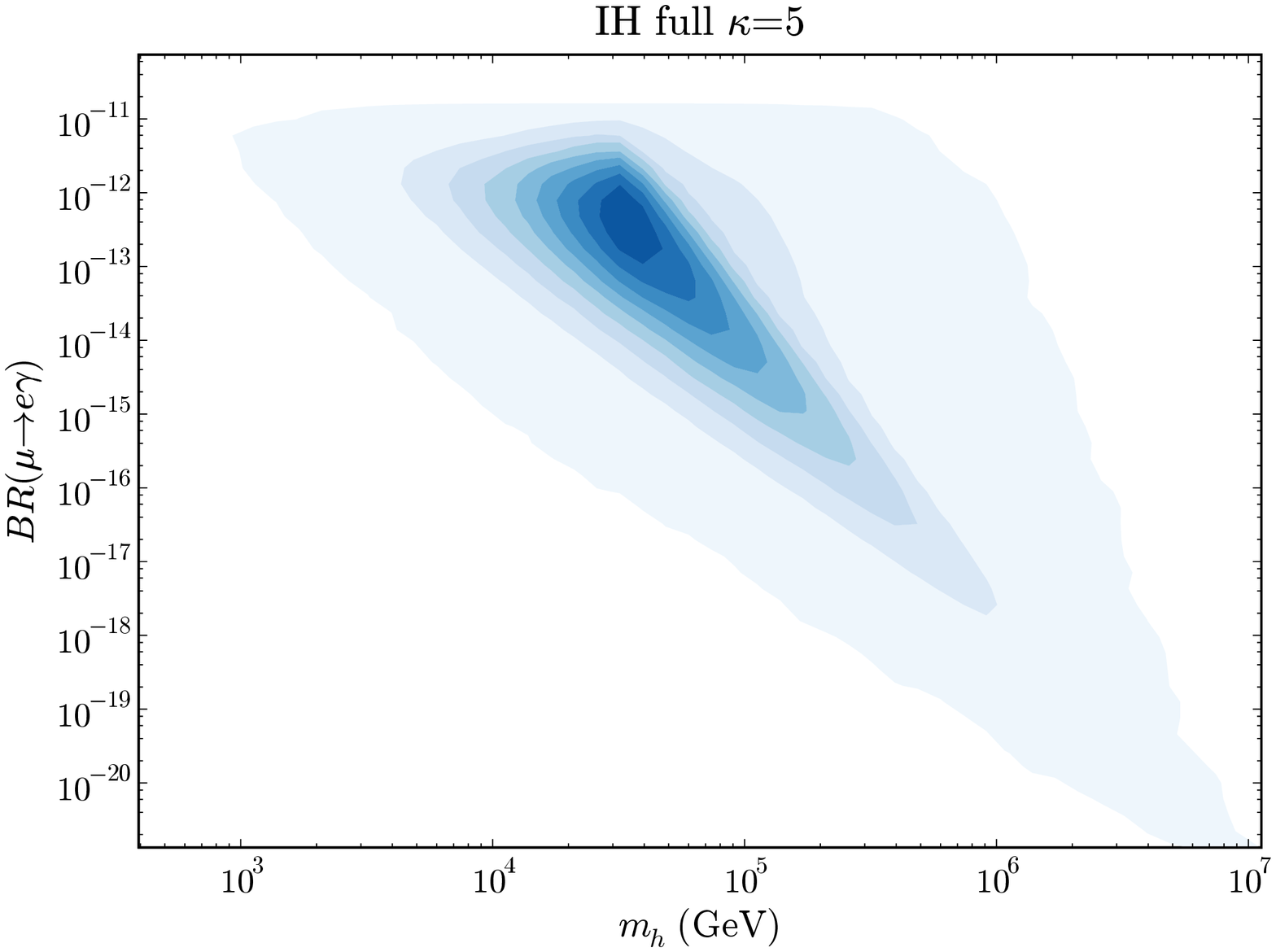}\includegraphics[width=0.49\columnwidth]{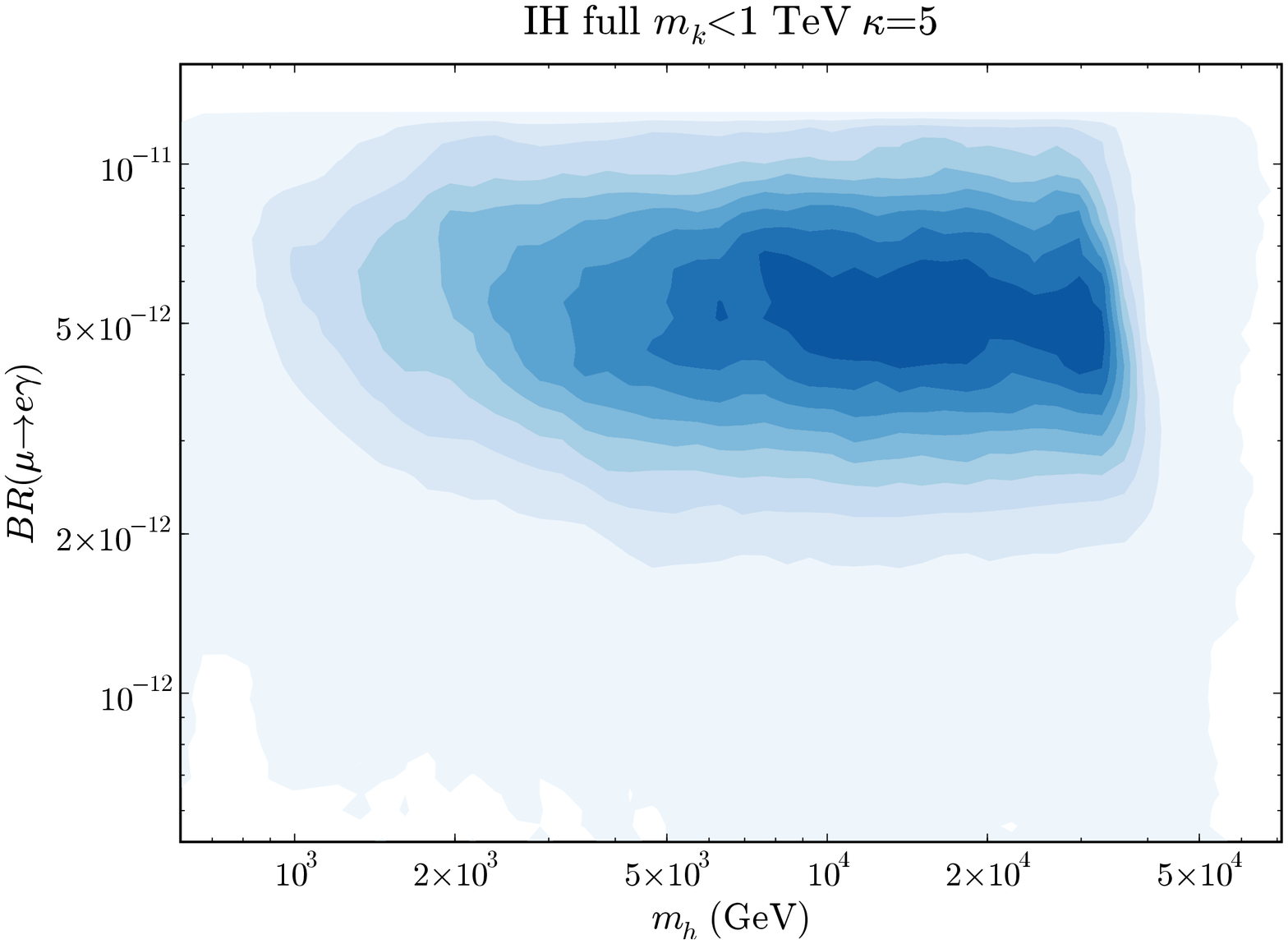}\par\end{centering}

\caption{\noindent IH: $BR(\mu\rightarrow e\gamma)$ vs $m_{h}$.\label{fig:IH-m2eg-mh}}
\end{figure}

\noindent Figure~\ref{fig:IH-m2eg-mh} is also similar to fig.~\ref{fig:NH:m2eg-mh}
but slightly more restrictive. In the general case we find $m_{h}\sim10^{2}-10^{7}\,\mathrm{GeV}$
and preferred values $m_{h}\sim40\,\mathrm{TeV}$. For $m_{k}<1\,\mathrm{TeV}$
the allowed range of $m_{h}$ is $m_{h}\sim500\,\mathrm{GeV}-70\,\mathrm{TeV}$
which will make its detection at the LHC problematic. 

\noindent The constraints on $BR(\tau\rightarrow3\mu)$ are also stronger
in the IH case, fig.~\ref{fig:IH-m2eg-mh}, than in the NH case. 
The allowed regions are similar
but more restrictive. Thus we find that in the general case the preferred
values of $BR(\tau\rightarrow3\mu)$ are in the $10^{-11}$ range,
although values as small as $10^{-21}$ are allowed. In the $m_{k}<1\,\mathrm{TeV}$
case the preferred values are $BR(\tau\rightarrow3\mu)\gtrsim 3\times 10^{-9}$
(to be compared with present limits $BR(\tau\rightarrow3\mu)<3.2\times10^{-8}$),
but values like $10^{-10}$ are not completely excluded. 

\begin{figure}[ht]
\noindent \begin{centering}\includegraphics[width=0.49\columnwidth]{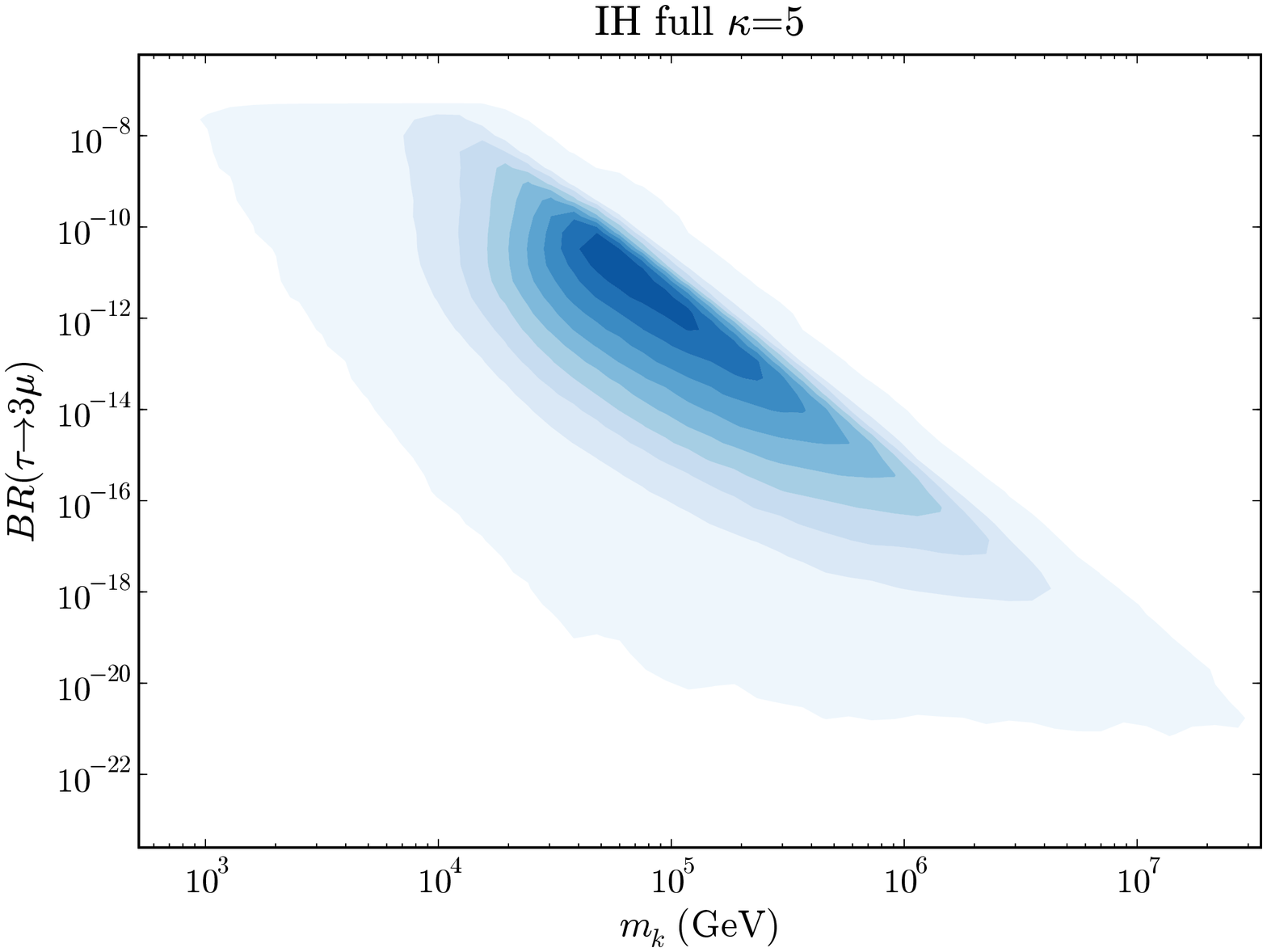}\includegraphics[width=0.49\columnwidth]{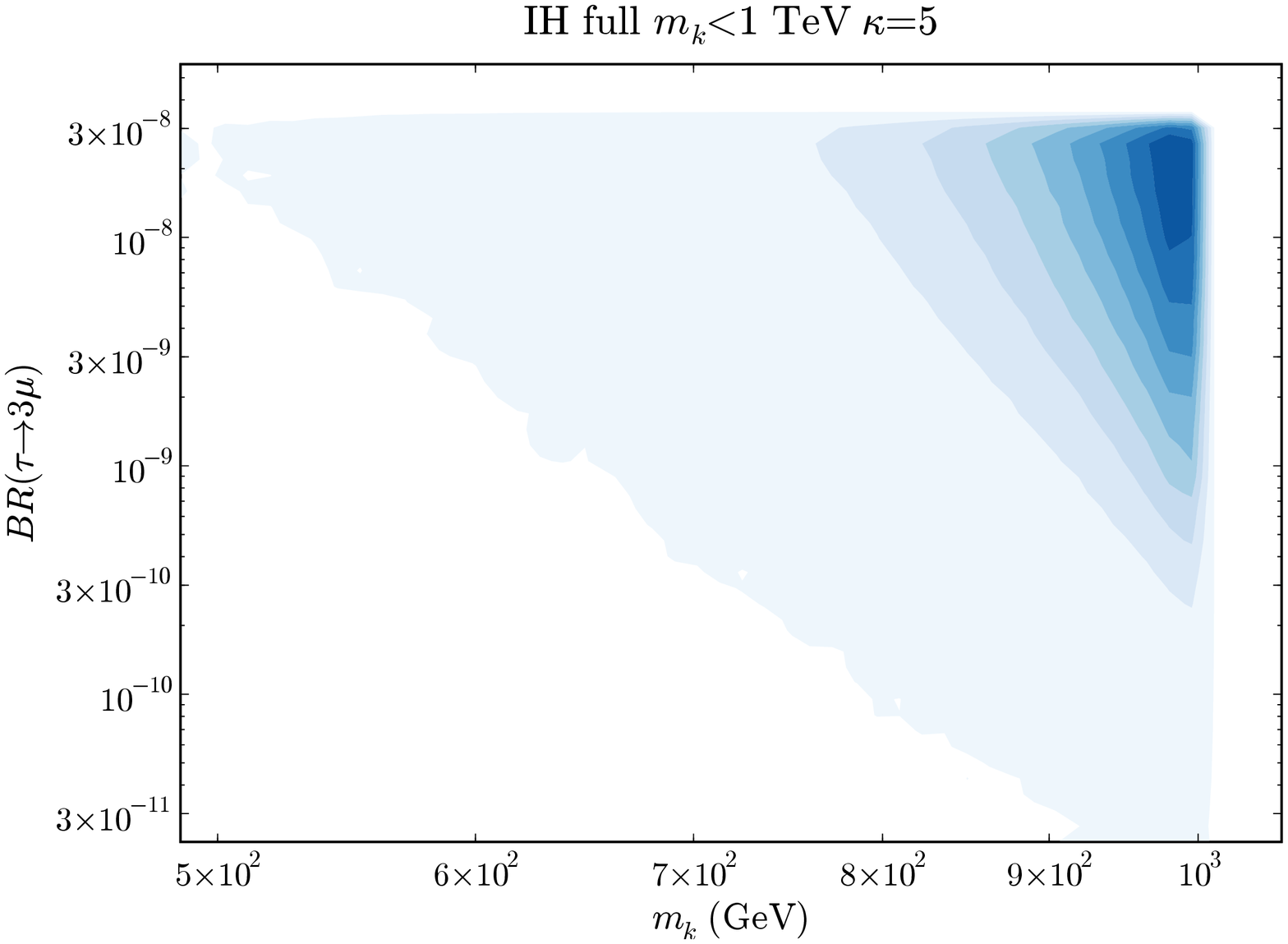}\par\end{centering}

\caption{\noindent IH: $BR(\tau\rightarrow3\mu)$ vs $m_{k}$.\label{fig:IH-t2mmm-mk}}
\end{figure}

\begin{figure}[ht]
\noindent \begin{centering}\includegraphics[width=0.49\columnwidth]{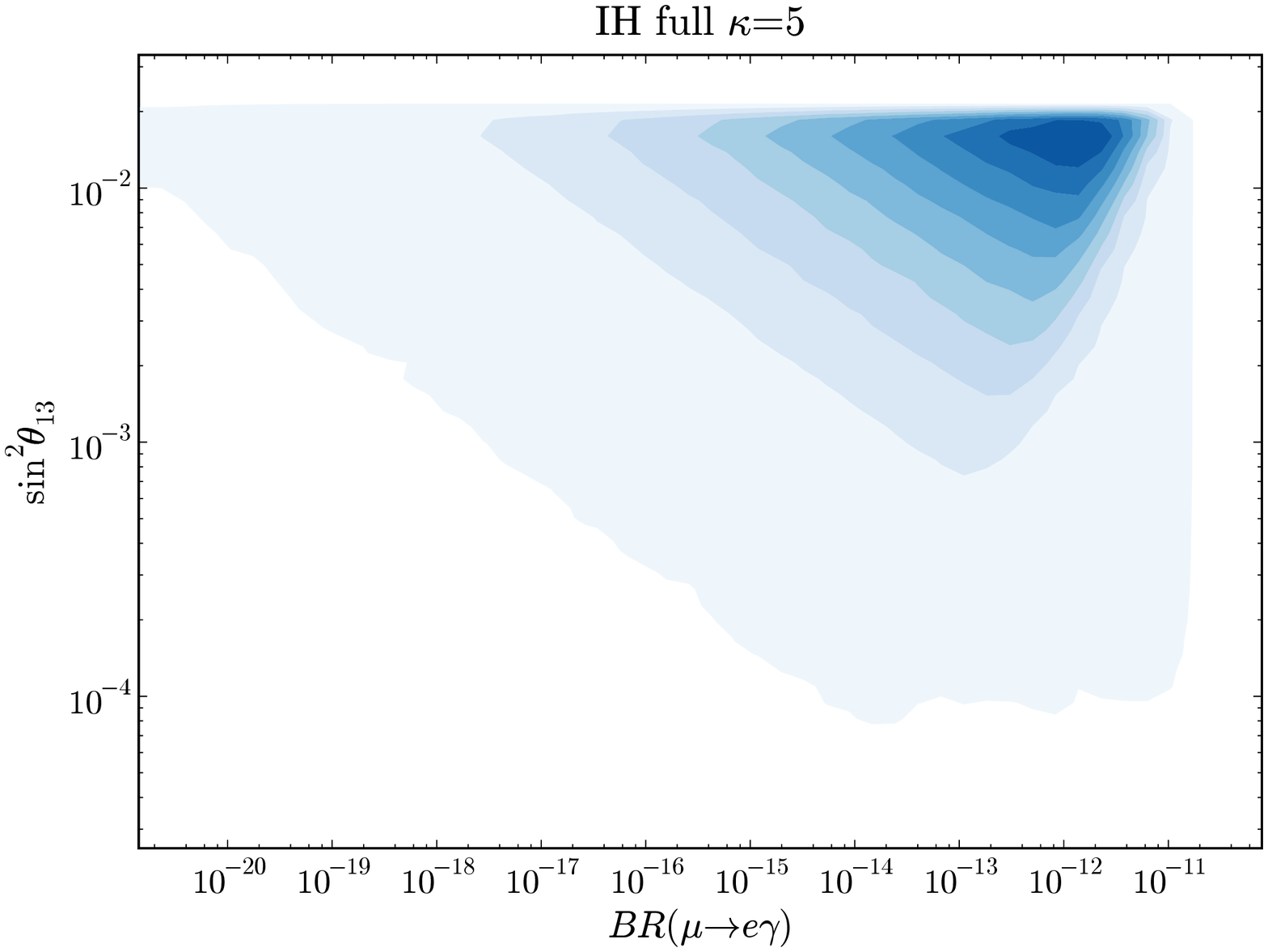}\includegraphics[width=0.49\columnwidth]{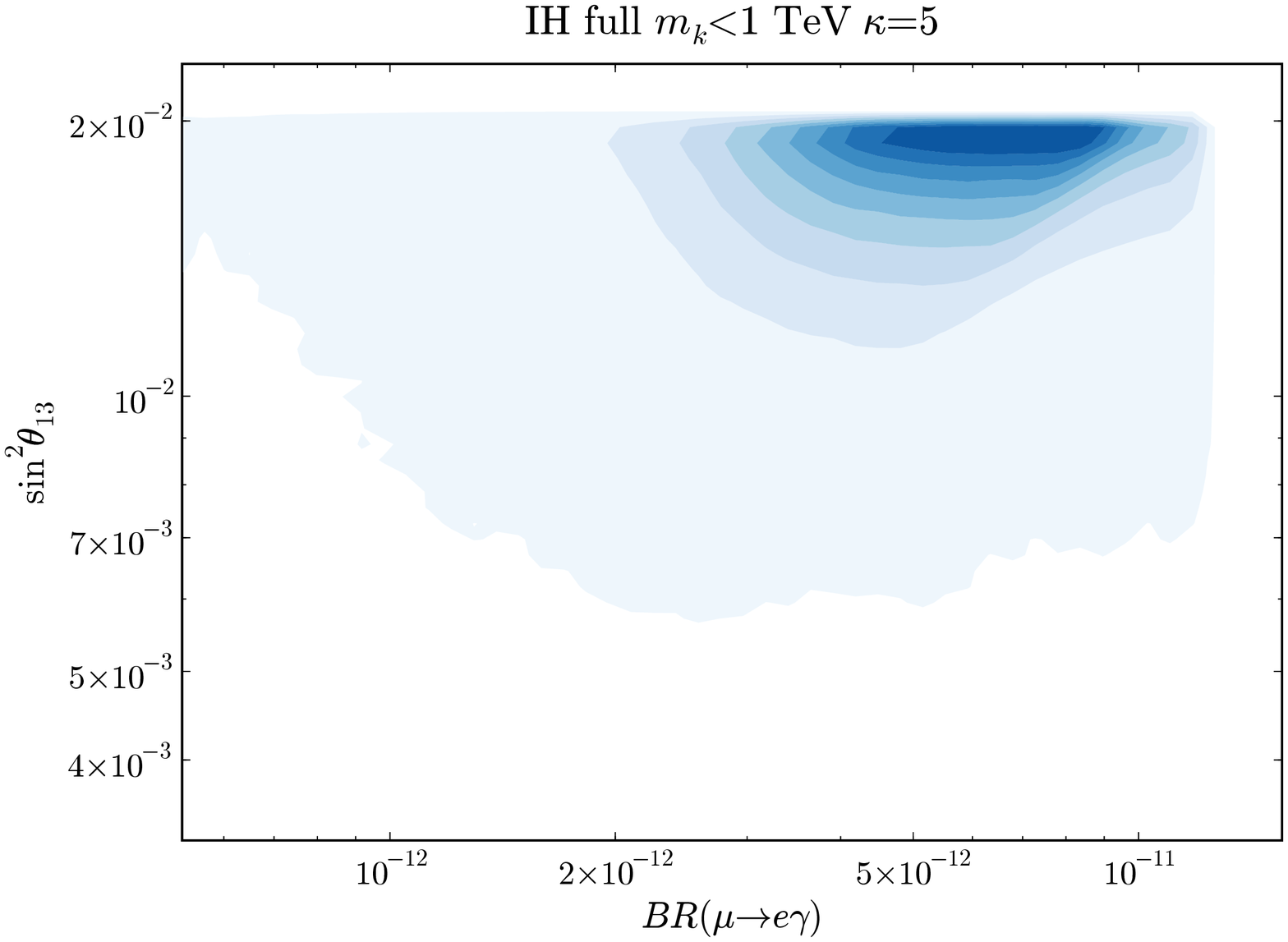}\par\end{centering}

\caption{\noindent IH: $\sin^{2}\theta_{13}$ vs $BR(\mu\rightarrow e\gamma)$.\label{fig:IH:s2th13-m2eg}}
\end{figure}

Finally, as has been shown analytically, in the IH case there is a
lower bound on $\sin^{2}\theta_{13}$. Thus in fig.~\ref{fig:IH:s2th13-m2eg}
we represent $\sin^{2}\theta_{13}$ versus $BR(\mu\rightarrow e\gamma)$.
We observe that there is a strong tendency towards relatively large
values of $\sin^{2}\theta_{13}$ even in the general case. The preferred
values are in the $\sin^{2}\theta_{13}\sim0.01$ region although values
below $10^{-4}$ do not seem completely excluded (the present upper
limit is $\sin^{2}\theta_{13}<0.02$). In addition smaller values
of $\sin^{2}\theta_{13}$ require larger values of $BR(\mu\rightarrow e\gamma)$.
If we also require that the $k^{++}$ can be discovered at the LHC
we find the preferred values of the model are constrained to a region
$\sin^{2}\theta_{13}\gtrsim0.01$ and $BR(\mu\rightarrow e\gamma)\gtrsim10^{-12}$.
We also see that values of $\sin^{2}\theta_{13}$ below $0.005$ and
$BR(\mu\rightarrow e\gamma)$ below $10^{-13}$ are very unlikely
in this case. Notice that mixings as small as $\sin^{2}\theta_{13}\sim0.005$
will be tested in a near future \cite{Huber:2006vr} and that the
MEG experiment, which will start this year, will probe $BR(\mu\rightarrow e\gamma)$
at the level of $10^{-13}$.

\noindent \begin{flushleft}\par\end{flushleft}

\section{Conclusions\label{sec:Conclusions}}

In this section we would like to highlight some of the most relevant
conclusions drawn from the previous analysis. 

The explanation of observed neutrino mixings in the model sets very
strong constraints on the structure of the couplings of the singly
charged scalars to fermions, in particular, in the NH case the couplings
must satisfy $f_{e\tau}\simeq f_{\mu\tau}/2\simeq f_{e\mu}$. To fix
the absolute value we need another observable, in the NH case we find
that the strongest bound comes from $\mu\rightarrow e\gamma$ and
tells us that $|f_{ei}|\lesssim0.05(m_{h}/\mathrm{TeV})$ and $|f_{\mu\tau}|\lesssim0.1(m_{h}/\mathrm{TeV})$.
In the case of IH the couplings must satisfy $f_{e\tau}\simeq-f_{e\mu}$
and $|f_{e\tau}|\gtrsim5|f_{\mu\tau}|$. Then, from lepton-hadron
universality we find $|f_{ei}|\lesssim0.1(m_{h}/\mathrm{TeV})$, and
$|f_{\mu\tau}|\lesssim0.02(m_{h}/\mathrm{TeV})$, bounds which are
similar to the bounds obtained from $\mu\rightarrow e\gamma$.

The structure of the couplings of the doubly charged scalar is also
very constrained by neutrino masses and mixings. In the case of NH
they must satisfy, to a good degree of precision, that $|g_{\mu\tau}|\approx|g_{\mu\mu}|(m_{\mu}/m_{\tau})$
and $|g_{\tau\tau}|\approx|g_{\mu\mu}|(m_{\mu}/m_{\tau})^{2}$. In
the case of IH this relation does not need to be satisfied exactly
because the electron couplings $g_{ei}$ can be relevant. However,
we have seen that in a large region of the parameter space this relation
is also required. Then, the best constraint comes from $\tau^{-}\rightarrow\mu^{+}\mu^{-}\mu^{-}$
which tells us that $|g_{\mu\mu}|\lesssim0.4(m_{k}/\mathrm{TeV)}$,
$|g_{\mu\tau}|\lesssim0.024(m_{k}/\mathrm{TeV)}$, $|g_{\tau\tau}|\lesssim0.0015(m_{k}/\mathrm{TeV)}$.
The $g_{ei}$ couplings are not constrained by neutrino data but are
constrained by low-energy processes which are summarized in tables~\ref{tab:meee}
and \ref{tab:meg}.

We find that the neutrinoless double beta decay parameter $\langle m_{\nu}\rangle_{ee}$
is strongly constrained in the model. We find $0.001\mbox{ \emph{\textrm{eV}}}<|\langle m_{\nu}\rangle_{ee}|<0.004\,\mathrm{eV}$
in the NH case and $0.01\mbox{ \emph{\textrm{eV}}}<|\langle m_{\nu}\rangle_{ee}|<0.06\,\mathrm{eV}$
in the IH case. This is just a consequence of the measured neutrino
masses and mixings and the particular structure of neutrino masses
of the model which predicts a massless neutrino.

If this model is the right explanation for neutrino masses and if
$m_{k}<\,1\,\mathrm{TeV}$ the LHC will produce more than $10$ events
in the $4lep$ channel (see section \ref{sec:LHC}) 
\footnote{We assume that, once efficiencies are taken into account, this corresponds
to $2$ reconstructed events \cite{Azuelos:2005uc,Gunion:1996pq}.%
}. There could be some dilution of the signal because the $k$ can
also decay into tau leptons or into two singly charged scalars but
in the case of NH this can only be relevant for $m_{k}>1\,\mathrm{TeV}$
or if the singly charged scalar is light enough, $m_{k}>2m_{h}$.
In the IH case the dilution of the $4lep$ signal is a bit larger,
still, most of the parameter space with $m_{k}<\,1\,\mathrm{TeV}$
will give more than $10$ events in the $4lep$ channel as long as the
$2h$ channel is not open.

If more than $10$ events are produced at the LHC in the $4lep$ channel
we find that $BR(\mu\rightarrow e\gamma)>10^{-13}$ in both the NH
and the IH cases. These values are precisely the sensitivity expected
in the MEG experiment at the Paul Scherrer Institute (PSI) which will
start to run soon \cite{Ritt:2006cg}. In fact one goal of the
experiment is to obtain a significant result before the start of the
LHC experiments. If this goal is achieved and nothing is seen we can
reverse the argument and claim that it will be very difficult to find
the charged scalars of this model at the LHC. We find that if
$BR(\mu\rightarrow e\gamma)<10^{-13}$, then, $m_{k}>900\,\mathrm{GeV}$
and $m_{h}>600\,\mathrm{GeV}$ in the NH case and that both scalar
masses will be above the $\mathrm{TeV}$ in the case of IH.

It is also important to remark that the photonic vector form factor
in muon-electron conversion in nuclei is enhanced with respect to
the tensor form factor due to logarithmic corrections of loops with
doubly charged scalars. Thus, if the current precision in muon-electron
conversion experiments is increased in the next years, there will
be additional tests on the model. 

If the doubly charged scalars are light enough to be produced at the
LHC there are also interesting contributions to rare tau decays. For
instance, in the IH case we find that most of the parameter space lies 
in the region $BR(\tau\rightarrow3\mu)\gtrsim 3\times 10^{-9}$, 
The NH case allows for slightly smaller branching ratios 
$BR(\tau\rightarrow3\mu)\gtrsim10^{-9}$. These  results have to
be compared with the the present limit, 
$BR(\tau\rightarrow3\mu)<3.2\times10^{-8}$ or the ranges that might
be explored in SuperB factories $\sim 10^{-9}$--$10^{-10}$
\cite{Akeroyd:2004mj,Bona:2007qt}.

The model gives a negative contribution to the $a_{\mu}=(g_{\mu}-2)/2$
of the muon. This means it cannot explain a positive deviation from
the SM. Conversely, the precise measurements of $a_{\mu}$ set interesting
constraints on the parameters of the model.

In general it is much more difficult to satisfy all the constraints
in the IH case than in the NH case. This can be seen in the MC acceptance
rate which is much lower in the IH case than in the NH case. In fact,
we have seen that satisfying all the neutrino mass data in the IH
case requires certain cancellations in the neutrino mass formulas
which imply that the neutrino Majorana phase $\phi$ cannot be zero,
actually, most of the parameter space lies in the region $e^{i\phi}\simeq-1$. 

Another interesting feature of the IH case is the emergence of a lower
bound on the $\theta_{13}$ mixing. We find that, even in the general
case, most of the allowed parameter space requires $\sin^{2}\theta_{13}\gtrsim0.001$,
although values below $10^{-4}$ do not seem completely excluded. In
addition, smaller values of $\sin^{2}\theta_{13}$ require larger values
of $BR(\mu\rightarrow e\gamma)$. If we also require that the $k^{++}$
is seen at the LHC through the $4lep$ channel we find that values
of $\sin^{2}\theta_{13}$ below $0.005$ are very unlikely (the present
upper limit is $\sin^{2}\theta_{13}<0.02$ and mixings as small as
$\sin^{2}\theta_{13}\sim0.005$ will be tested in a near future \cite{Huber:2006vr}
). 

In short, the requirement that the model is able to explain the observed
pattern of neutrino masses and mixings places very strong limits on
the parameters of the model. Thus, if the doubly charged scalar of
the model is seen at the LHC through the $4lep$ channel, the model
predicts large contributions to several low energy processes, 
$\mu\rightarrow e\gamma$,
$\tau\rightarrow3\mu$, hadron-lepton universality tests, which should
be within reach of the next round of experiments. In particular, the
MEG, $\mu\rightarrow e\gamma$ experiment \cite{Ritt:2006cg} should
provide results at the required level of precision before the start
of the LHC experiments. If the $k^{++}$ is discovered at the LHC
and MEG does not see anything, the model will be in serious trouble.
In all the other cases the model can fit all the data. Moreover,
if the $k^{++}$ is discovered at the LHC and MEG
sees $\mu \rightarrow e \gamma$ in the allowed region, the model will
be a serious candidate to explain neutrino masses.

\begin{acknowledgments}
This work has been supported in part by the MEC (Spain) under the
grant FPA2005-00711 and by the EU MRTN-CT-2006-035482 (FLAVIAnet).
M.N. acknowledges financial support from {}``Fundaç\~ao para a Ciência
e a Tecnologia'' (FCT, Portugal) through the projects PDCT/FP/63912/2005,
PDCT/FP/63914/2005, CFTP-FCT UNIT 777, and by the EU MRTN-CT-2006-035505
(HEPtools).
\end{acknowledgments}

\end{document}